\documentclass[review]{elsarticle}

\usepackage{lineno,hyperref}
\usepackage{amsmath}
\usepackage{pgfplots}
\usepackage{xcolor}

\begin{document}

\begin{frontmatter}

\title{A data-constrained sharp Immersed Boundary Method for aerospace applications}


\author[Poitiers,RTech]{M.A. Chemak}
\author[RTech]{E. Constant}
\author[Lille]{M. Meldi}
\cortext[Lille]{Corresponding author, \textit{marcello.meldi@ensam.eu}}
\address[Poitiers]{Institut Pprime, Department of Fluid Flow, Heat Transfer and Combustion, CNRS -
ENSMA - Universit\'{e} de Poitiers, UPR 3346, SP2MI - T\'{e}l\'{e}port, 211 Bd. Marie et Pierre Curie,
B.P. 30179 F86962 Futuroscope Chasseneuil Cedex, France}
\address[RTech]{RTech, Verniolle, France}
\address[Lille]{Arts et Métiers ParisTech, CNRS, Univ. Lille, ONERA, Centrale Lille, UMR 9014- LMFL- Laboratoire de Mécanique des fluides de Lille- Kampé de Feriet, F-59000 Lille, France}

\begin{abstract}
A numerical tool relying on sharp Immersed Boundary Method (IBM) is developed for the analysis of aerospace applications. The method, which is conceived for application using segregated solvers relying on implicit time discretization, uses a Luenberger observer to dynamically update the free coefficients governing the numerical algorithm. This technique improves the accuracy of the method and permits to target the representation of complex flow features at the wall, taking into account the velocity field and heat transfer. The method is used to investigate several test cases of increasing complexity, including a space vehicle during atmospheric reentry. The tool exhibits interesting efficacy in terms of accuracy versus computational costs required.  
\end{abstract}


\end{frontmatter}


\section{Introduction}

The accurate prediction of flow features for aerospace applications requires the development of advanced numerical tools able to precisely account for complex features of the flow over a large range of regimes. The latter may include large variations of the Reynolds number $Re$ and the Mach number $Ma$. In fact, for applications such as atmospheric reentry, flow configurations from hypersonic $Ma$ down to subsonic $Ma$ are observed over a single trajectory \cite{HAYARAMOS201639}. Different problematic aspects associated with heat exchange, chemical reactions and surface erosion interact and exhibit complex features through the dynamic evolution of the flow. Therefore, the development of numerical methods to provide accurate prediction of a large range of physical configurations is a key challenge for such studies. Improvement in the representation of dynamic conditions using Computational Fluid Dynamics (CFD) is critical for the identification and control of reentry conditions, which may result in catastrophic accidents if not accurately predicted. 


Among the different proposals in the literature, the Immersed Boundary Method (IBM) \cite{Peskin1972_jcp,Mittal2005_arfm,deTullio_2007,Kim2019_ijhff,Verzicco2022_arfm} is a family of tools aiming to represent the flow features around immersed bodies using non-conforming meshes. In this case, the presence of the body is not taken into account by an explicit internal boundary condition, like in classical body-fitted approaches. Instead, manipulation of the dynamic equations via addition of source term and / or imposing local features is used to mimic the presence of the immersed body. The main advantages of this family of techniques are that i) thanks to the use of regular grids, the discretization error is under control and ii) updates of the position of the body in time due to its movement can be taken into account without expensive re-meshing procedures. This last point is essential for example in the case of surface erosion or icing, which are commonly observed phenomena in aerospace engineering. The main disadvantage of IBM tools when compared with body-fitted representation is the relative lower precision at similar grid resolution. This problematic aspect is usually magnified at high $Ma$ regimes where several concurring physical mechanisms can be observed.   
Recent advances on this aspect by Peron et al. \cite{peron2021immersed} rely on a sharp-interface IBM method which provides excellent performance for the representation of compressible flows. This Ghost-cell method uses mirror points i.e. imposing a prescribed physical behavior in the mesh elements in the solid domain of the body, so that the targeted behavior at the wall is obtained. Such a strategy appears to be significantly more precise than classical diffused-interface techniques which generally underperform for supersonic regimes \cite{Kim2019_ijhff}. This sharp-interface IBM procedure, which is tailored for explicit flow solvers, is more difficult to implement in implicit codes which are extensively used in industrial analysis.

In the present work, we propose an  IBM technique based on the work of Peron et al. \cite{peron2021immersed}. The research advancement here proposed is that the method is adapted to both explicit and implicit solvers. The main difference is that the condition in the solid region is not obtained imposing values for the flow field, but a source term is instead included in the dynamic equations. This source term is optimized via a data-driven procedure based on a Luenberger observer \cite{doi:10.1080/00207179308934406} which optimizes the intensity of the source terms included. A procedure is also implemented to provide an adaptive modification of the gain of the observer in time, in the case of analysis of unstationary flows. The method developed is integrated within the opensource C++ code \textit{OpenFOAM}. The solvers from this available module have been extensively used in the literature for academic and industrial analyses \cite{Tabor2010553,Selma2014241,Margheri2014_cf,Zhang2020_cf}, which also include applications of diffused-interface IBM \cite{Constant2017_cf,Riahi2018_jcp,Riahi2024}.  The methodology is assessed for the analysis of compressible flows accounting for heat exchange at the body surface.


The structure of the article is now presented. In Section \ref{sec::num} the numerical tools used in the present analysis are described. These include the CFD solvers used as well as the reference IBM technique considered. In Section \ref{sec::IBMLun} the proposed IBM is introduced. Detailed explanation is provided for its integration within the numerical solver as well as for the dynamic data-driven procedure to optimize it. In Section \ref{sec::Results} the performance of the method is validated. In Section \ref{sec::sphere} the method is used to investigate the flow around a sphere, while an application to complex geometries for aerospace applications is presented in Section \ref{sec::vehicle}. Finally, in Section \ref{sec:Conclusions} are drawn.

\section{Numerical tools}
\label{sec::num}

The numerical ingredients employed in the present analysis are described in this Section. These tools include the numerical solvers for the simulation of the flow, the algorithm for the Immersed Boundary Method and the data-driven strategy used to augment the latter.

\subsection{Numerical solvers}

The numerical platform used to perform the flow simulations is \textit{OpenFOAM}. This C++ library, which has been extensively used for academic studies as well as industrial applications, provides Finite Volume discretization of the Navier--Stokes equations. Several resolution algorithms taken from the open literature \cite{Ferziger1996_springer} are proposed, which are tailored to the flow condition investigated. In the present work two of these solvers are used, which are developed for incompressible and compressible flows, respectively.    

\subsubsection{Incompressible flow solver: PisoFoam}

The first solver used, \textbf{PisoFoam}, is a transient semi-explicit pressure based code. The PISO algorithm \cite{ISSA198640} is used to predict and correct the velocity field $\mathbf{u}$ for incompressible flows. The differential equations to be discretized, which are here written for Newtonian fluids, are the momentum equation and the Poisson equation:
\begin{eqnarray}
    \frac{\partial \mathbf{u}}{\partial t} + \nabla \cdot (\mathbf{u} \mathbf{u}) - \nabla \cdot (\nu \nabla \mathbf{u}) = - \frac{1}{\rho} \nabla p +\mathbf{f}_\mathbf{u} \label{eq:MomInc} \\
    \frac{1}{\rho} \nabla^2 p= - \nabla \cdot (\nabla \cdot (\mathbf{u} \mathbf{u})) + \nabla \cdot \mathbf{f}_\mathbf{u} \label{eq:PoisInc}
\end{eqnarray}

where $\mathbf{u}$ is the velocity, $p$ is the pressure, $\rho$ is the density, $\nu$ is the kinematic viscosity and $\mathbf{f}_\mathbf{u}$ is a volume forcing term. The latter can be employed to enforce the IBM effect in the physical domain. The numerical resolution of equations \ref{eq:MomInc} - \ref{eq:PoisInc} is based on a number of sequential operations. For each time step, the following loop is iteratively repeated until a prescribed convergence is reached:

\begin{enumerate}
\item Resolution of the discretized momentum equation \label{eq:MomInc}:

\begin{equation}
a_P \mathbf{u}_P = -\sum_N a_N \mathbf{u}_N + \Phi_0 (\mathbf{u}_0) - \frac{1}{\rho} \nabla p +\mathbf{f}_\mathbf{u} = \Phi (\mathbf{u}) - \frac{1}{\rho} \nabla p +\mathbf{f}_\mathbf{u} \label{incompressible_momentum}
\end{equation}

where the term $\Phi_0$ is the result of the discretization of the time derivative, which relies on the velocity field $\mathbf{u}_0$ at previous time steps. The subscripts $P$ and $N$ refer to the velocity field calculated in the center of the mesh element considered and its neighbors, respectively. Also, the coefficients $a_P$ and $a_N$ are determined by the discretization schemes used to approximate the spatial derivatives of the velocity field. Therefore, the operator $\Phi$ includes the representation of the non-linear interactions of equation \ref{eq:MomInc}. At the very first iteration, $\nabla p$ is determined using the pressure field from the previous time step.

\item The resolution of equation \ref{incompressible_momentum} does not grant a solenoidal condition of the velocity field, which is a necessary condition for incompressible flows. Using the assumption that the discrepancy is due to the gradient of the pressure field, the latter is re-calculated using the new velocity field via the discretized Poisson equation:
\begin{equation}
\frac{1}{\rho} \nabla \cdot \left( \frac{\nabla p}{a_P} \right) = \nabla \cdot \left( \frac{\Phi (\mathbf{u})+ \mathbf{f}_\mathbf{u}}{a_P} \right) \label{incompressible_Pois}
\end{equation}
Equations \ref{incompressible_momentum} - \ref{incompressible_Pois} are then iteratively resolved until convergence is reached.
\end{enumerate}

\subsubsection{Compressible flow solver: RhoCentralFoam}

The second solver used for compressible flow simulations is \textbf{RhoCentralFoam}.
It is a transient semi-explicit density based solver. It solves the
following mass, momentum and energy equations:

\begin{equation}
\frac{\partial\rho}{\partial t}+\nabla\left(\rho\mathbf{u}\right)=f_{\rho}
\label{compressible_mass}
\end{equation}

\begin{equation}
\frac{\partial\left(\rho\mathbf{u}\right)}{\partial t}+\nabla\left(\rho\mathbf{u}\mathbf{u}\right)=-\nabla p+\nabla\left(\tau\right)+\rho\mathbf{f}_{\mathbf{u}}
\label{compressible_momentum}
\end{equation}

\begin{equation}
\frac{\partial\left(\rho e_{t}\right)}{\partial t}+\nabla\left(\rho e_{t}\mathbf{u}\right)=-\nabla\left(p\mathbf{u}\right)+\nabla\left(\tau\mathbf{u}\right)-\nabla q+\rho\mathbf{u}.\mathbf{f}_{\mathbf{u}}+\rho c_{v}f_{T}
\label{compressible_energy}
\end{equation}

where $\tau$ is the viscous stress tensor,
$e_{t}$ is the total energy, $q$ is the the diffusive heat flux expressed
as $q=\lambda\nabla T$, $c_{v}$ is the specific heat capacity at
constant volume, and $f_{\rho}$, $\mathbf{f}_{\mathbf{u}}$ and $f_{T}$
are respectively volume forcing terms in the mass, momentum and energy
equations. It is important to observe that the term $f_{\rho}$ appears to be violating the mass conservation of the flow. However, it is known that numerical / modelling errors in IBM can be responsible of mass flux through the body surface \cite{Verzicco2022_arfm}, and the term $f_{\rho}$ (which is small in magnitude) is used to compensate such numerical problem. At each time step, the governing equations are solved
sequentially:
\begin{enumerate}
\item Resolution of the mass equation. The density field is updated as:

\[
a_{\rho}\rho^{n+1}=\phi_{\rho}\left(\rho^{n},\mathbf{u}^{n}\right)+f_{\rho}^{n}
\]

where $a_{\rho}$ represents the time discretization coefficient,
and $\phi_{\rho}$ includes the convective flux, as well as part of
the time derivative discretisation.
\item The momentum equation is resolved a first time without diffusive flux and
without forcing term, i.e. the inviscid compressible momentum equation:

\begin{equation}
\frac{\partial\left(\rho\mathbf{u}\right)}{\partial t}+\nabla\left(\rho\mathbf{u}\mathbf{u}\right)=-\nabla p^{n}
\end{equation}

At this step the conservative variable $\left(\rho\mathbf{u}\right)$
is solved and the intermediate value $\left(\rho\mathbf{u}\right)^{*}$
can be written as:

\begin{equation}
a_{\mathbf{u}}\left(\rho\mathbf{u}\right)^{*}=\phi_{\mathbf{u}}\left(\left(\rho\mathbf{u}\right)^{n},\mathbf{u}^{n}\right)-\nabla p^{n}
\end{equation}

where $a_{\mathbf{u}}$ represents the time discretization coefficient,
and $\phi_{\mathbf{u}}$ includes the convective flux of the momentum
equation, as well as part of the time derivative discretisation.
\item An intermediate velocity field is computed as $\mathbf{u^{*}}=\left(\rho\mathbf{u}\right)^{*}/\rho^{n+1}$,
then it used for the second resolution of the momentum equation including
the diffusive flux and the momentum forcing term:

\begin{equation}
a_{\mathbf{u}}\rho^{n+1}\mathbf{u}^{n+1}=\phi_{\mathbf{u}}\left(\left(\rho\mathbf{u}\right)^{n},\mathbf{u}^{n}\right)+\phi_{\mathbf{u}}^{'}\left(\mathbf{u^{*}}\right)-\nabla p^{n}+\rho^{n+1}\mathbf{f}_{\mathbf{u}}^{n}
\end{equation}

where $\phi_{\mathbf{u}}^{'}$ includes the diffusive flux. The momentum
is updated as $\left(\rho\mathbf{u}\right)^{n+1}=\rho^{n+1}\mathbf{u}^{n+1}$.
\item The total energy equation is solved first without diffusive heat flux
and without forcing term:

\begin{equation}
\frac{\partial\left(\rho e_{t}\right)}{\partial t}+\nabla\left(\rho e_{t}\mathbf{u}\right)=-\nabla\left(p\mathbf{u}\right)+\nabla\left(\tau\mathbf{u}\right)
\end{equation}

At this step the conservative total energy is solved:

\begin{equation}
a_{e}\left(\rho e_{t}\right)^{*}=\phi_{e}\left(\left(\rho e_{t}\right)^{n},\mathbf{u}^{n}\right)-\nabla\left(p^{n}\mathbf{u}^{n+1}\right)+{\color{red}\nabla\left(\tau\mathbf{u}\right)}
\end{equation}

where $a_{e}$ represents the time discretization coefficient, and
$\phi_{e}$ includes the convective flux of the total energy, as well
as part of the time derivative discretization.
\item An intermediate sensible energy is computed as $e_{s}^{*}=\left(\rho e_{t}\right)^{*}/\rho^{n+1}-0.5\mathbf{u}^{n+1}.\mathbf{u}^{n+1}$,
then the sensible energy equation including the diffusive heat flux
and the energy forcing term. The term $\rho\mathbf{u}.\mathbf{f}_{\mathbf{u}}$
associated with the work produced by $\mathbf{f}_{\mathbf{u}}$ is neglected, because it is very small in magnitude and it is usually responsible of fluctuations that can affect the stability of the calculation.

\begin{equation}
a_{e}\rho^{n+1}e_{s}^{n+1}=\phi_{e}\left(\left(\rho e_{t}\right)^{n},\mathbf{u}^{n}\right)-\nabla\left(p^{n}\mathbf{u}^{n+1}\right)+{\color{red}\nabla\left(\tau\mathbf{u}\right)}-\nabla\left(q\left(e_{s}^{*}\right)\right)+\rho^{n+1}c_{v}f_{T}^{n}
\end{equation}
The temperature field is updated using $e_{s}^{n+1}$, \textbf{$\left(\rho e_{t}\right)^{n+1}=\rho^{n+1}\left(e_{s}^{n+1}+0.5\mathbf{u}^{n+1}.\mathbf{u}^{n+1}\right)$},
and the pressure is update using the perfect gas law.
\end{enumerate}

\section{Sharp IBM based on a Luenberger observer}
\label{sec::IBMLun}

\begin{figure}
\begin{centering}
\includegraphics[scale=1]{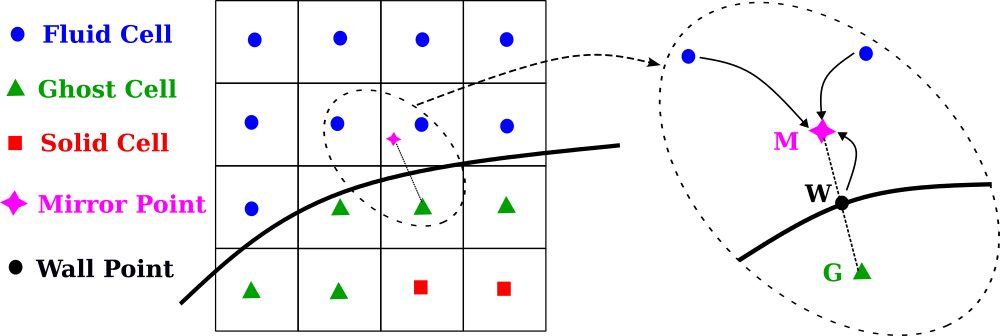}
\par\end{centering}
\caption{Scheme of application of the ghost cell method proposed by Peron et al. \cite{peron2021immersed}.}
\label{fig:ghostCells}

\end{figure}

The classical implementation of sharp IBM for flow simulation \cite{FADLUN200035,Mohd-Yusof97} relies on ghost cells which are selected in the solid region behind the immersed boundary. Explicit treatment of the flow field in such cells in performed to obtain a suitable behavior in the fluid region. The IB technique developed by Peron et al. \cite{peron2021immersed} is articulated through the following steps (see figure \ref{fig:ghostCells} for a qualitative representation):
\begin{enumerate}
\item Ghost cells are identified via the position of the surface of the immersed body. These cells may change in time if the body is moving.
\item For each center G of a ghost cell, a neighbor point W is identified over the surface of the immersed body. This point is the projection of G over the body surface.
\item A third point M is identified, for which the relation $\overrightarrow{GW}=\overrightarrow{WM}$ is satisfied. If M does not coincide with the center of a grid element, the physical quantities ($\mathbf{u}$,$p$,$T$...) are interpolated using information from the neighbor cell centers
\item The flow field in G is determined exploiting symmetric / antisymmetric behavior of the solution over the segment GWM and using the information available on M and the boundary condition to be applied in W (wall condition). For the latter, the two most classical options are imposing a value for the variable investigated (Dirichelet condition) or for its gradient (Neumann condition). For a general variable $\alpha$, if one wants to obtain a Dirichelet condition for $\alpha(W)=a$, then the value on the ghost cell is selected so that $\alpha(G)=a -\alpha(M)$. Similarly, a Neumann boundary condition can be obtained imposing a field value on G, after observation of the same variable on M. 
\end{enumerate}

This sharp IBM has proven accurate and robust for the analysis of compressible flows \cite{peron2021immersed}. However, the formulation previously proposed is adapted for the inclusion in explicit flow solvers. Extrapolation towards implicit solvers, where recursive procedures need to be performed, may be complex due to variations in the variables in the correspondence of the ghost cells. Implicit solvers are implemented in numerous open source codes in the fluid mechanics community, because of their robustness, memory management and relaxed constraints in terms of time steps. Therefore, the extension of the present methodology to a larger platform of solvers will be beneficial for analyses of industrial test cases including moving immersed bodies with complex geometries.

In this work, we propose an advancement of the method by Peron et al. \cite{peron2021immersed} which is compatible with explicit and implicit flow solvers. The method replicates the same structure previously presented in points 1-3 and illustrated in figure \ref{fig:ghostCells}. However, the boundary conditions are not obtained imposing a value for the physical quantities on the ghost cells. A source term is derived instead and integrated in the dynamic equations in correspondence of the ghost cells, which are now resolved. The source term is adapted using a Luenberger observer \cite{doi:10.1080/00207179308934406}, which can take into account variations in time of the physical quantities. This methodology is extremely flexible and it can be tailored to the features of the solver employed, exhibiting a high potential of portability between different numerical codes. In the specific case of OpenFOAM, the strategy employed for incompressible or compressible solvers is different, owing to the algorithms used in time advancement of the solution. Details about such implementation are provided in the following.

\subsection{Implementation of the source terms and boundary condition used}

The integration of the methodology presented in the solvers of the platform OpenFOAM is now detailed. The source terms considered are the following:
\begin{enumerate}
    \item $\mathbf{f}_u$ for the momentum equation \ref{incompressible_momentum} for incompressible solvers
    \item $\mathbf{f}_\rho$,$\mathbf{f}_u$ and $\mathbf{f}_T$ and  for the mass, momentum and energy equations respectively \ref{compressible_mass} ,\ref{compressible_momentum} \ref{compressible_energy} for compressible solvers
\end{enumerate}

The volume source terms are considered as vector / scalar quantities defined on the whole physical domain, but their value is set to zero outside the ghost cells. The calculation of their value, which is determined separately for each ghost cell, is now explained. First, a target value for the physical quantities of interest is determined using the same strategy presented in the work by Peron et al. \cite{peron2021immersed}, which was summarized in Section \ref{sec::IBMLun}. This target value is the one calculated using the information (in terms of flow field and boundary conditions) from the points M and W in figure \ref{fig:ghostCells}. However, now this target value is not imposed anymore on the ghost cell, but it is used to drive the performance of a Luenberger observer which determines the value of the volume forcing. Thanks to the observer, the intensity of the source term is proportional to the discrepancy between the flow field calculated by the CFD solver on the ghost cell and the target value. Let us consider a flow variable $\alpha$ on the ghost cell G (see figure \ref{fig:ghostCells}). The initial estimate of the volume forcing is calculated as:
\begin{equation}
\mathbf{f}_{\alpha}= \mathbf{f}_{0} \, \frac{E}{a+\left|E\right|} \; , \qquad E=\frac{\alpha^{G,target}-\alpha^{G}}{\left|\alpha^{G,target}\right|} 
\label{eq:forc-lun}
\end{equation}
where $\mathbf{f}_{0}$ is a prescribed initial value, $E$ is the normalized discrepancy between the variable $\alpha$ calculated on the ghost cell by the CFD solver and the value targeted, and $a$ is an empiric coefficient driving the stiffness of the gain of the Luenberger observer. The forcing proposed in equation \ref{eq:forc-lun} may exhibit very high values in case of large discrepancy between the calculated and targeted solution, which is usually observed at the very beginning of the numerical simulation. This issue can produce very large positive-negative oscillations during the first time steps, leading to divergence of the calculation. This effect can be reduced increasing the value of the coefficient $a$, but this strategy usually precludes optimized performance once the discrepancy diminishes. In order to effeciently palliate this issue, the time evolution of the source terms is controlled via an algorithm that accounts for the information at the previous time step. Considering advancement from the instant $n-1$ to $n$, the proposed final algorithm is the following:
\begin{equation}
\mathbf{f}_{\alpha}^{n}= \mathbf{f}_{\alpha}^{n-1} + \left|\mathbf{f}_{\alpha}^{n-1} \right| \frac{E}{a+\left|E\right|}  
\label{eq:forc-lun-corr}
\end{equation}
this implementation allows for a continuous variation in time of the volume forcing terms, which greatly improves the stability of the calculation in particular during the initial time steps. 

The rationale behind the choice of the target values for each source term is now discussed. The source term $\mathbf{f}_u$ targets the non-slip condition for the velocity field in correspondance of the wall i.e. $\mathbf{u}(W)=\mathbf{0}$. In order to impose such boundary condition, the target velocity vector is defined using properties of antisymmetry of the field:
\begin{equation}
    \mathbf{u}^{G,target}=-\mathbf{u}^{M}
\end{equation}
This condition is imposed for both incompressible and compressible solvers. For the latter, more constraints are imposed to take into account the behavior of the other variables at the wall. In the case of an adiabatic boundary condition, the target temperature in the ghost cell is simply equal to the interpolated temperature in the mirror point:
\begin{equation}
T^{G,target}=T^{M}
\end{equation}
In the case of imposed temperature at the wall $T^W$, the target temperature is obtained via a linear interpolation formula, using information from the points W and M:
\begin{equation}
T^{G,target}=2T^{W}-T^{M}
\end{equation}
Finally, the pressure $p$ and the density $\rho$ are discussed. In order to ensure a zero pressure-gradient at the wall, the condition $p^{G,target}= p^{M}$ must be fulfilled. Using the perfect gas law, the target density can be expressed as:
\begin{equation}
\rho^{G,target}=\frac{\rho^{M}T^{M}}{T^{G,target}}
\end{equation}
Once $\rho^{G,target}$ is computed, a source term $f_\rho$ is determined to be included in the mass equation. It is important to stress that, while this term may in theory lead to violation of the mass conservation, this effect is almost negligible for every test case investigated. In fact, the magnitude of this term is low and one must consider that it is always applied to the ghost cell i.e. in a solid region.

\section{Validation of the IBM tool}
\label{sec::Results}

The IBM presented in the previous section is validated for the simulation of the two dimensional NACA 0012 airfoil in an incompressible laminar flow. This academic test case is a difficult one for IBM, because of its slender shape and the presence of a sharp edge. The results will be compared to the DNS results of Kurtulus \cite{Kurtulus}. The sensitivity of the results to the mesh resolution and size of the time step will be assessed via the generation of a numerical database. All the simulations are performed for a Reynolds number based on freestream velocity $U_\infty$ and chord length $c$ which is $Re_c = U_\infty c / \nu = 1000$. The angle of attack chosen for the investigation is 11°. The flow at these conditions is unsteady, which will also allow to evaluate the suggested IBM implementation in dynamic conditions.

\subsection{Case setup}
\begin{figure}
\begin{centering}
\includegraphics[scale=1.0]{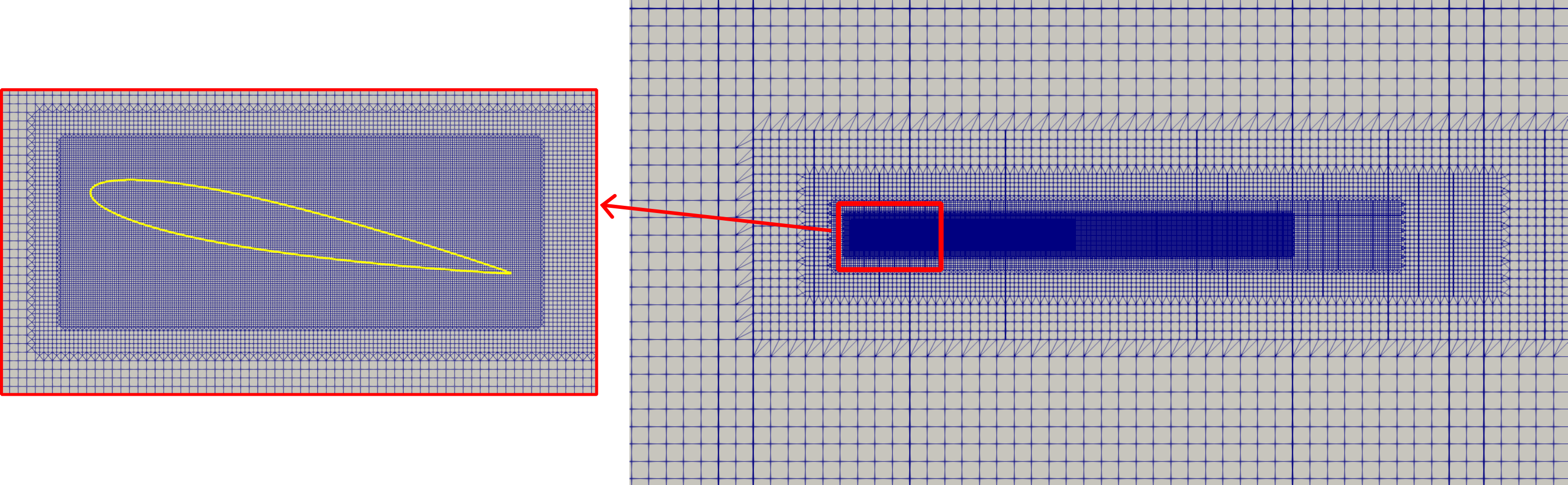}
\par\end{centering}
\caption{\label{fig:mesh_profile}Mesh grid topology used for the simulation of NACA 0012 airfoil. The yellow line represent the immersed boundary}
\end{figure}

The computational domain is rectangular and with dimensions of $20\,c$ in the streamwise direction and $12\,c$ in the vertical direction. The leading edge of the airfoil is placed at distance equal to $6\,c$ from the inflow. The chord line is first aligned with the grid then the immersed airfoil is rotated to have an angle of attack of 11°. The base grid is Cartesian and the mesh elements have a square shape. Grid refinement boxes were applied around the airfoil and in the wake region as shown in Fig. \ref{fig:mesh_profile}. Table \ref{tab:naca_cases} summarizes the performed simulations and their corresponding mesh (G) and time step (T) characteristics. Runs were performed with three different grids \textbf{G1}, \textbf{G2} and \textbf{G3} with cell sizes of $c/200$, $c/400$ and $c/800$ in the proximity of the immersed body, respectively. The sensitivity of the results to the time step was assessed on mesh grid \textbf{G2} with three different time steps. {These time steps are constants and equal to $T1 = 4 \cdot 10^{-4} t_A$, $T2 = 2 \cdot 10^{-4} t_A$ and $T3 = 1 \cdot 10^{-4} t_A$, where $t_A = c / U_\infty$ is the characteristic advection time.} Finally, a last simulation is performed using the grid \textbf{G3bis}, which is based on \textbf{G3} with an additional refinement level applied on the leading and trailing edges. {For this simulation, the time step is equal to $T4= 2 \cdot 10^{-5} t_A$.} All the cases were initialized with uniform pressure field equal to $p_{\infty}$, and a uniform velocity field $\mathbf{u}=[0.5 U_\infty, \, 0, \, 0]$. The permanent regime is reached for approximately  $t=80 t_A$, which minor differences observed for the different grids used. Once the permanent regime is reached and the initial conditions are dissipated, the results are averaged in time for $t=20 t_A$ to calculate the bulk flow statistics and the time-averaged flow fields.

\begin{table}
\begin{centering}
\begin{tabular}{|c|c|c|c|c|}
\hline 
Case & $\Delta x=\Delta y$ & Edge refinement & number of cells & $\Delta t \cdot t_A$ \tabularnewline
\hline 
\hline 
\texttt{profile\_G1\_T2} & $c/200$ & not applied & 72k & $2 \cdot 10^{-4}$\tabularnewline
\hline 
\texttt{profile\_G2\_T1} & $c/400$ & not applied & 125k & $4 \cdot 10^{-4}$\tabularnewline
\hline 
\texttt{profile\_G2\_T2} & $c/400$ & not applied & 125k & $2 \cdot 10^{-4}$\tabularnewline
\hline 
\texttt{profile\_G2\_T3} & $c/400$ & not applied & 125k & $1 \cdot 10^{-4}$\tabularnewline
\hline 
\texttt{profile\_G3\_T2} & $c/800$ & not applied & 300k & $2 \cdot 10^{-4}$\tabularnewline
\hline 
\texttt{profile\_G3Bis\_T4} & $c/800$ & applied & 438k & $2 \cdot 10^{-5}$\tabularnewline
\hline 
\end{tabular}
\par\end{centering}
\caption{\label{tab:naca_cases} Characteristics of the database of numerical simulations for the 2D NACA0012 airfoil. Details for the grid and the time step are provided.}
\end{table}

\subsection{Results and discussion}

\begin{figure}
\begin{centering}
\includegraphics[scale=0.2]{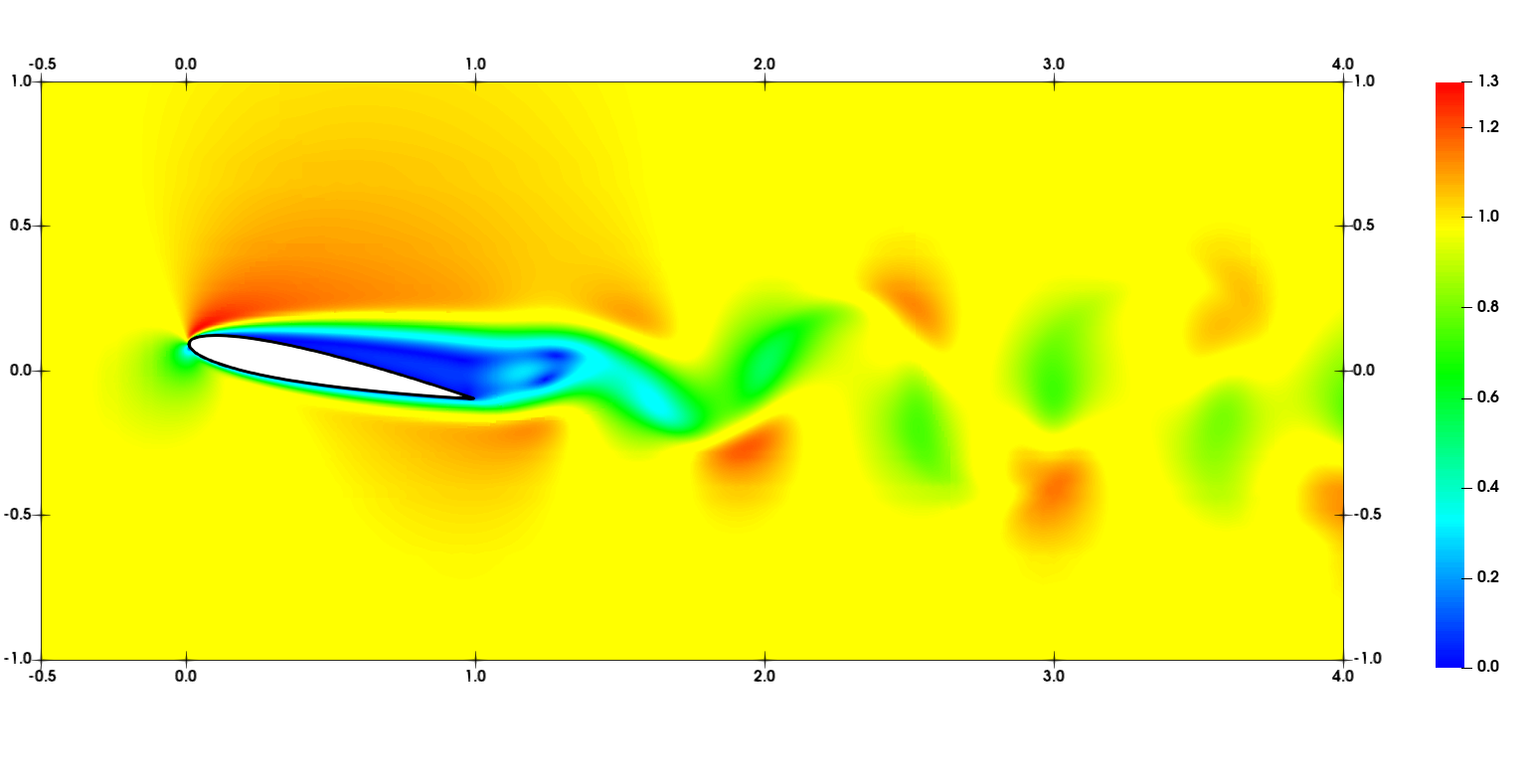}
\par\end{centering}
\caption{\label{fig:instant_vel_naca}Instantaneous velocity magnitude field calculated for the case \texttt{profile\_G3Bis\_T4}}
\end{figure}

\begin{figure}
\begin{tabular}{cc}
\includegraphics[scale=0.15]{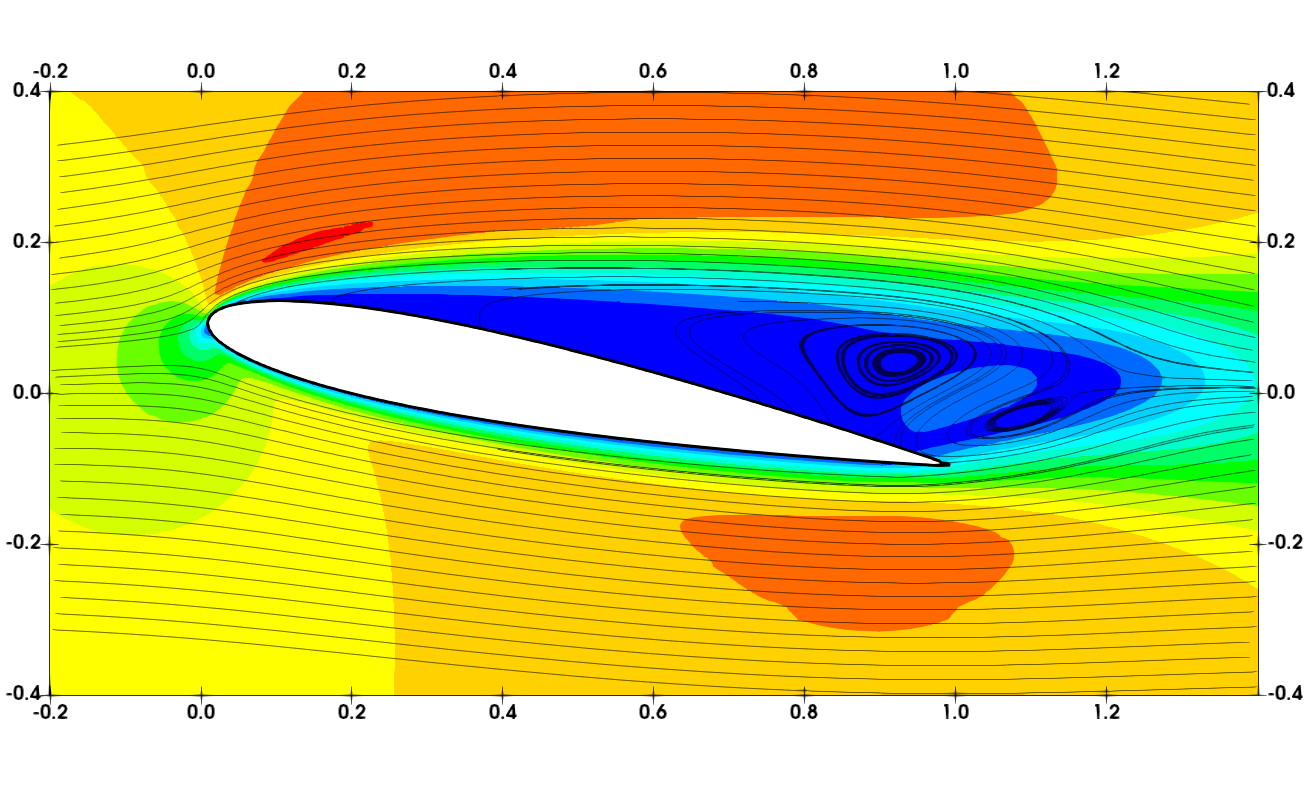} & \includegraphics[scale=0.15]{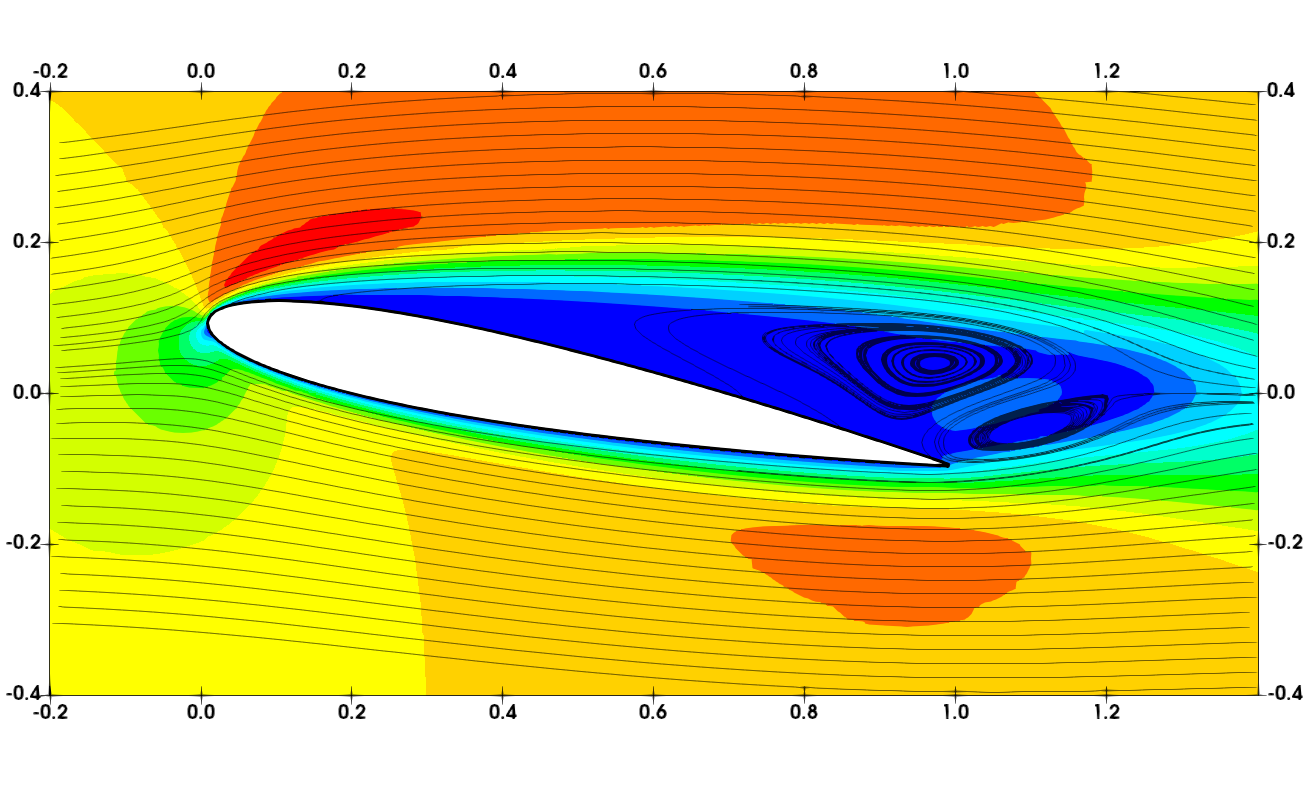} \\
(a) & (b) \\
\includegraphics[scale=0.15]{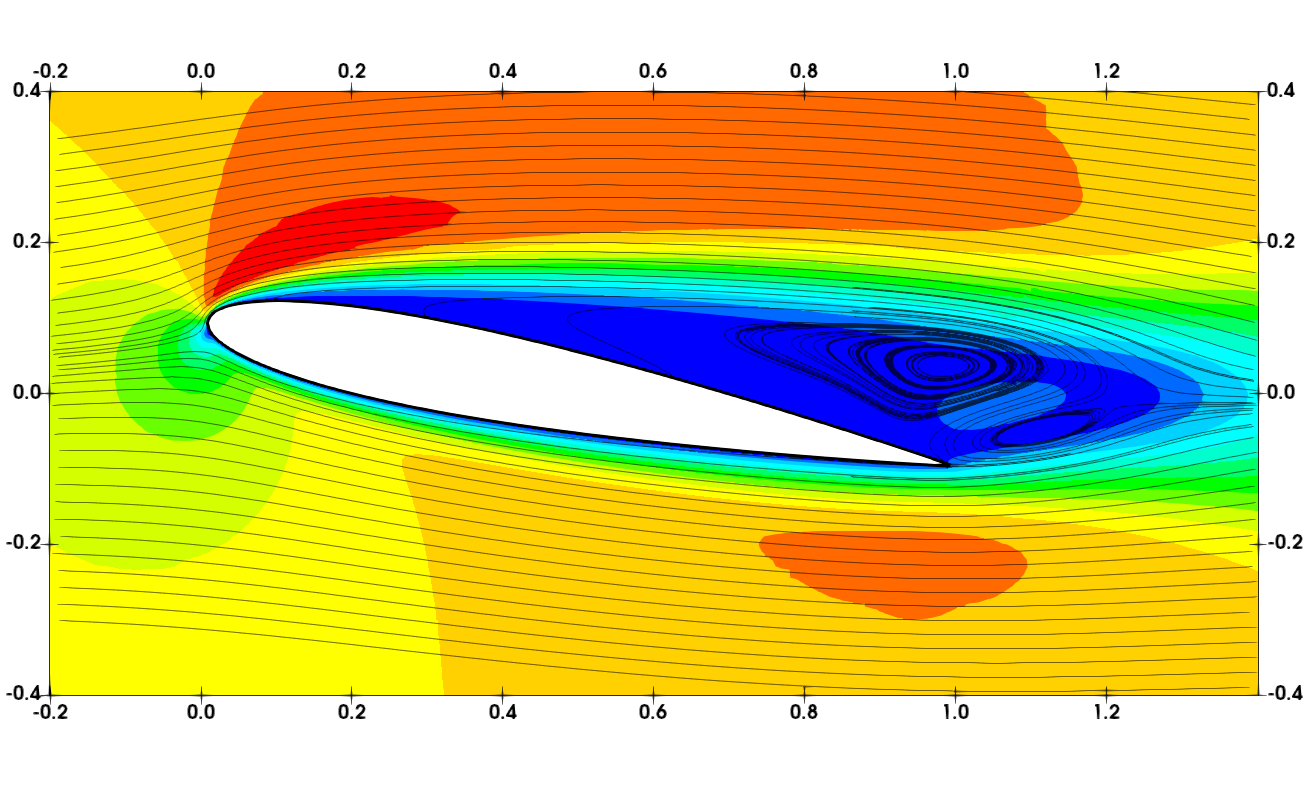} & \includegraphics[scale=0.15]{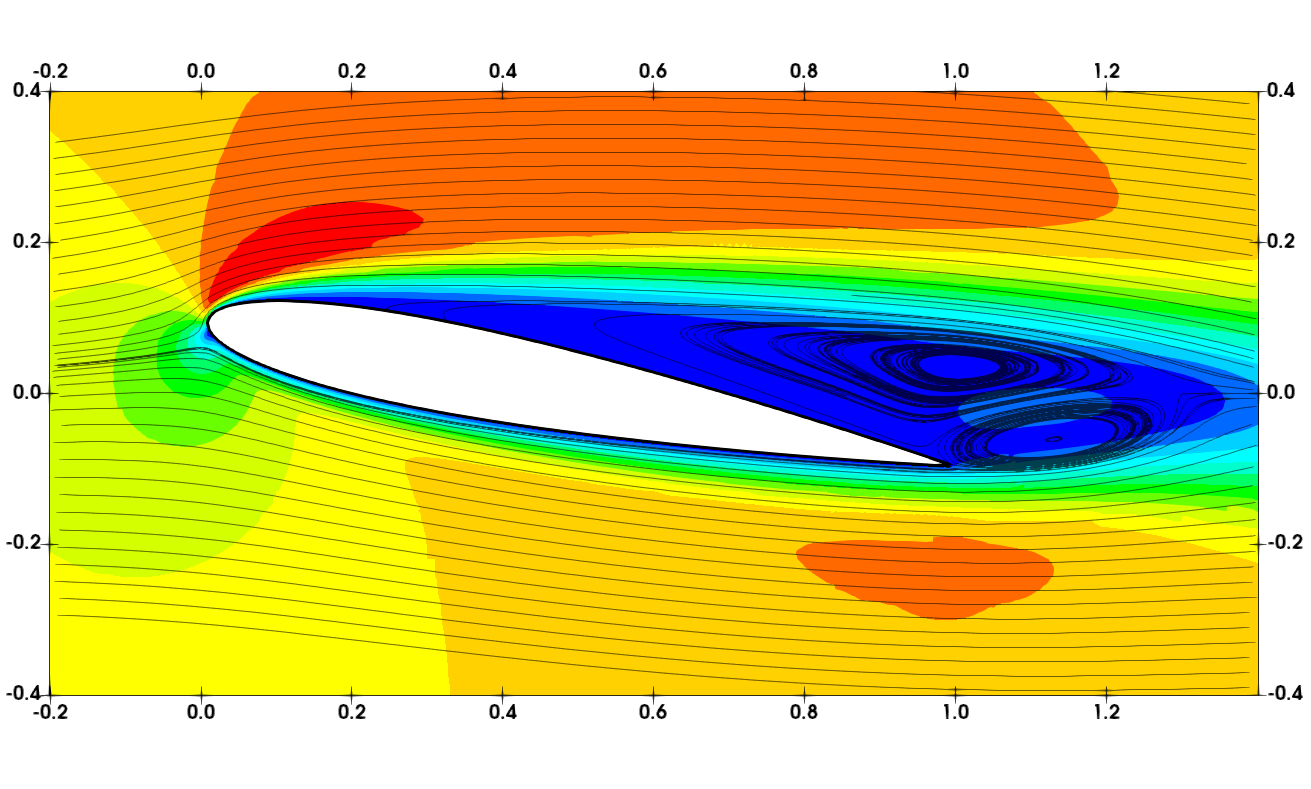} \\
(c) & (d)
\end{tabular}
\caption{\label{fig:average_vel_field} Comparison of the average velocity magnitude field for four simulation of the database. Black lines represent velocity streamlines. Results are shown for the simulations (a) \texttt{profile\_G1\_T2}, (b) \texttt{profile\_G2\_T2}, (c) \texttt{profile\_G2\_T3} and (d) \texttt{profile\_G3Bis\_T4} }
\end{figure}

\begin{figure}
\begin{centering}
\includegraphics[scale=0.5]{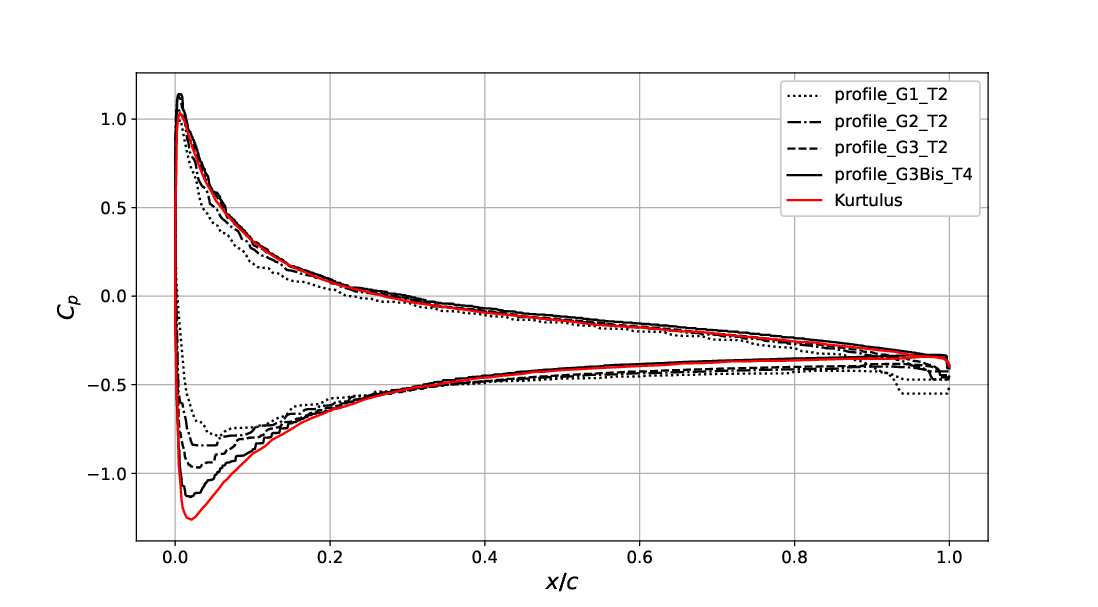}
\par\end{centering}
\caption{\label{fig:cp_profile}Mean pressure distribution around the profile NACA 0012 obtained via IBM. Results are compared with data from a reference body-fitted simulation.}
\end{figure}

First of all, the analysis of the results indicates that the proposed IBM is able to capture the physics of the flow investigated. A qualitative visualization of the unsteady flow is shown in figure \ref{fig:instant_vel_naca}, where one can see that the flow topology is well represented. The fields are shown for the simulation of the case \texttt{profile\_G3Bis\_T4}, but every runs capture the macroscopic features of the flow. Some differences can be observed via a fine analysis of the database. The time-averaged velocity magnitude $\overline{u_M}$ is shown in Fig. \ref{fig:average_vel_field} for four runs performed. All the case predict similar pattern of streamlines of average velocity close to the trailing edge, which are also similar to the ones reported by Kurtulus \cite{Kurtulus} (see Fig. 8) for the same angle of attack. Figs. \ref{fig:average_vel_field} (a) and \ref{fig:average_vel_field} (b) use the same time step, but the grid is significantly more refined. A stronger acceleration of the flow at the leading edge is here observed for the latter, indicating therefore a mild sensitivity of the results to the mesh used. Comparisons of simulations shown in Figs. \ref{fig:average_vel_field} (b) and \ref{fig:average_vel_field} (c) highlights the sensitivity to the size of the time step. For the choices here performed, the sensitivity of the flow field is very weak. Finally, the comparison of \ref{fig:average_vel_field} (b) and \ref{fig:average_vel_field} (d), where the medium and very refined grids are used, indicates that the latter predicts a slightly larger recirculation region, but the flow fields are very similar. One first conclusion that can be drawn is that the method appears to produce robust results even with coarse grids, and in particular the procedure is numerically stable. The distribution of the mean pressure coefficient $C_p$ around the airfoil is now investigated. Fig. \ref{fig:cp_profile} shows this profile for different grids and results are compared with body-fitted simulations reported in the literature \cite{Kurtulus}, which are highlighted in red. The comparison of the results indicates that a global good agreement is observed, even if differences are locally observed. In particular, the largest discrepancies with the body fitted simulation aare observed at the upper surface in the proximity of the leading edge, where the acceleration of the flow is maximum. One can see that differences between the usage of grids G2 and G3 are relatively small, indicating again that the model obtains sufficiently accurate results even with a relatively coarse grid.

The accuracy of the method is now quantitatively investigated comparing the calculated bulk flow quantities. To do so, a L1 $(a_{IBM}-a_{ref})/a_{ref}$ norm is employed, where $a$ is a quantity of interest. The subscript $IBM$ indicates a quantity calculated using the IBM method, while the subscript $ref$ indicates results from the body-fitted reference simulation. Table \ref{tab:naca_results} presents the average lift and drag coefficients, $C_L$ and $C_D$, for all the cases and their relative error to the reference results by \cite{Kurtulus}. The pressure and viscous contributions of the drag, $C_{D,p}$ and $C_{D,v}$ respectively, are also reported and their relative error is computed with respect to the total drag coefficient. The relative error observed for the lift coefficient exhibits a large variability for the database of IBM runs performed, from $19\%$ down to $1.5\%$. A similar trend is obtained for the drag coefficient, with variations going from $38.5\%$ to $6\%$. However, when the drag coefficient is decomposed in pressure ($C_{D,p}$) and viscous ($C_{D,v}$) part, one can see that the error for the former is relatively low (lower than $10\%$) even for the most coarse grids used. On the other hand, the error on the drag coefficient due to viscous phenomena is extremely sensitive to grid resolution. This is expected as $C_{D,v}$ explicitly depends on the calculated velocity gradients, whose prediction degrades fast with grid coarsening. Also, the IBM source term do not include any information about velocity gradients, so that the methodology is not able to actively control this physical quantity.
The Strouhal number based on the instantaneous lift oscillations is also calculated. For all the cases with the exception of case \texttt{profile\_G3Bis\_T4}, a discrepancy of approximately 5\% is observed, which is weakly dependent on the grid resolution and the time step.

In conclusion, the results obtained from the database of the simulations here performed indicate that an acceptable level of accuracy is obtained by the proposed IBM technique. In particular, a robust prediction of the drag coefficient associated with pressure was obtained, with a global discrepancy lower than $10\%$ even for the coarsest grids used. This favorable feature is of interest for preliminary analyses for atmospheric reentry, where massive flow separation is observed and the pressure contribution to the drag coefficient usually is close to $90-95\%$.   

\begin{table}
\begin{centering}
\begin{tabular}{|c|c|c|c|c|c|c|c|c|c|c|}
\hline 
Case & $St$ &  & $C_{L}$ &  & $C_{D}$ &  & $C_{D,p}$ &  & $C_{D,v}$ & \tabularnewline
\hline 
\hline 
Reference & 0.825 &  & 0.460 &  & 0.182 &  & 0.107 &  & 0.075 & \tabularnewline
\hline 
\texttt{profile\_G1\_T2} & 0.881 & 5.51\% & 0.373 & -18.91\% & 0.251 & 38.5\% & 0.123 & 8.79\% & 0.128 & 29.7\%\tabularnewline
\hline 
\texttt{profile\_G2\_T1} & 0.877 & 5.03\% & 0.405 & -12.0\% & 0.230 & 26.4\% & 0.125 & 9.89\% & 0.105 & 16.5\%\tabularnewline
\hline 
\texttt{profile\_G2\_T2} & 0.874 & 4.67\% & 0.416 & -9.57\% & 0.225 & 23.6\% & 0.125 & 9.89\% & 0.100 & 13.7\%\tabularnewline
\hline 
\texttt{profile\_G2\_T3} & 0.858 & 4.0\% & 0.415 & -9.78\% & 0.215 & 18.1\% & 0.121 & 7.69\% & 0.094 & 10.4\%\tabularnewline
\hline 
\texttt{profile\_G3\_T2} & 0.881 & 5.51\% & 0.442 & -3.91\% & 0.211 & 15.9\% & 0.123 & 8.79\% & 0.088 & 7.14\%\tabularnewline
\hline 
\texttt{profile\_G3Bis\_T4} & 0.847 & 2.67\% & 0.453 & -1.52\% & 0.193 & 6.04\% & 0.113 & 3.29\% & 0.08 & 2.74\%\tabularnewline
\hline 
\end{tabular}
\par\end{centering}
\caption{\label{tab:naca_results}Strouhal number and aerodynamic coefficients with relative errors for the NACA 0012 simulations}
\end{table}

\section{Supersonic flow around a sphere}
\label{sec::sphere}

\begin{figure}
\begin{centering}
\includegraphics[scale=0.2]{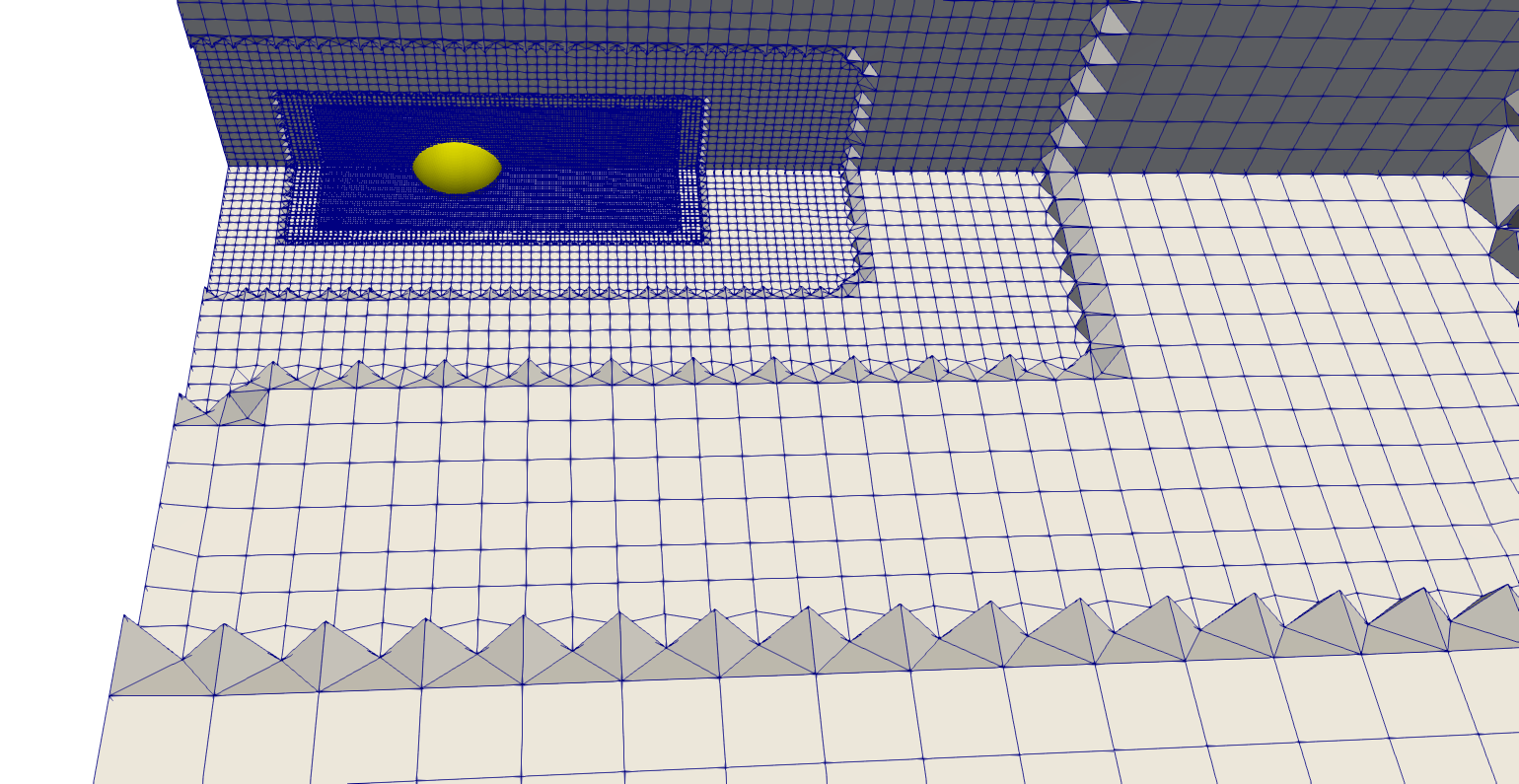}
\par\end{centering}
\caption{Qualitative representation of the grid used to perform the calculations of the supersonic flow around a sphere. The grid shown in figure is G1.}
\label{fig:meshSphere}
\end{figure}

The academic case of supersonic flow around a sphere is now investigated using the proposed IBM. This three-dimensional test case has been selected because of its increased complexity in terms of flow features as well as to study heat transfer phenomena using the IBM. These thermal effects have been studied imposing i) adiabatic conditions and ii) constant temperature on the surface of the immersed sphere. The mesh topology used for the calculations is shown in figure \ref{fig:meshSphere}. This strategy for the distribution of the grid elements has been chosen following criteria used in the work by Riahi et al. \cite{Riahi2018_jcp} and in particular the center of the sphere is set on the origin of the axes $x,y,z$. The boundary conditions are now discussed. At the inlet, a uniform fixed value is set for the velocity $U_\infty$ in the $x$ direction, the pressure $p_\infty$ and the temperature $T_\infty$. Zero-gradient boundary conditions are set for the physical quantities in the lateral surfaces and at the outlet. For this test case, $x$ is the streamwise direction and $y$ and $z$ are the normal direction. The performance of the solver has been tested analyzing its sensitivity to the following parameters:
\begin{itemize}
\item Grid refinement. Three different grids namely G1, G2 and G3 have been used to perform the simulations. The mesh distribution is very similar for the three grids and it is the one shown in figure \ref{fig:meshSphere}. The main difference is in this case represented by the resolution employed in the sphere region. A region is  identified  within the cylinder whose main axis is aligned with the $x$ direction, it has extremes at $x=-1.5D$ and $x=2.5D$ and a radius of $1.5D$. Within this cylinder, which includes the sphere, special requirements for grid refinement are imposed. For G1, $\Delta x= \Delta y = \Delta z = D/20$, where $D$ is the diameter of the sphere. This resolution is increased to $D/40$ for the grid G2 and $D/80$ for G3, respectively. Outside of the box indicated, the grid used is almost identical. {The total number of grid elements for G1, G2 and G3 is $3.5 \cdot 10^5$, $10^6$ and $3.4 \cdot 10^6$, respectively.} 
\item Mach number $Ma$. Two inflow Mach numbers are considered, namely $Ma=1.2$ and $Ma=3$.
\item Reynolds number $Re$. Three values of the inflow $Re= \rho U_\infty D / \mu$ are considered, $Re=100$, $Re=200$, and $Re=300$. For the values of $Ma$ and $Re$ investigated, the flow around the sphere is laminar and stationary.  
\item Thermal surface condition of the sphere. Four conditions are tested. The first one is the adiabatic condition i.e. zero-gradient behavior of the temperature in the proximity of the wall. In addition, a constant temperature wall condition has been investigated. Several values of the parameter $TR= T_{SS} / T_\infty$ have been investigated, where $T_{SS}$ is the temperature imposed on the surface of the sphere and $T_\infty$ is the inflow temperature. The case of a heated sphere is here considered and values selected for this parameter are $TR=1.1$, $TR=1.5$ and $TR=2.0$.
\end{itemize}

The body-fitted Direct Numerical Simulations performed by Nagata and coworkers \cite{nagata_direct_2018,nagata_nonomura_takahashi_fukuda_2020,Nagata2016} are used as reference to validate the accuracy and robustness of the method. The characteristics of the database of simulations run for the analysis of this test case are summarized in Tab. \ref{tab::DataBSphere}.

\subsection{Results and discussion}

\begin{figure}
\begin{tabular}{cc}
\includegraphics[scale=0.35]{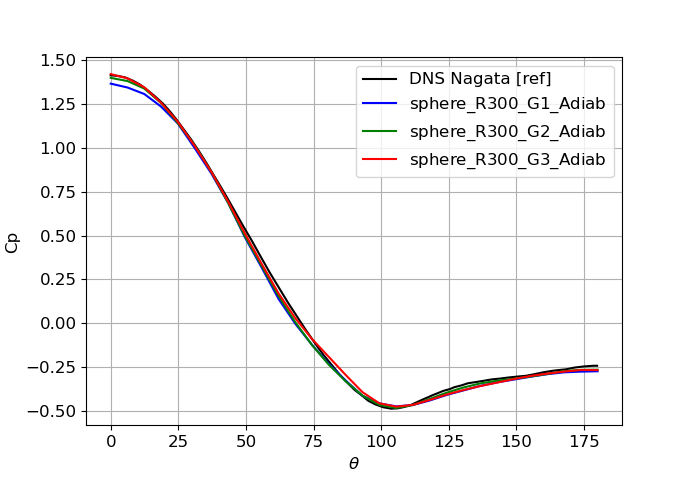} & \includegraphics[scale=0.35]{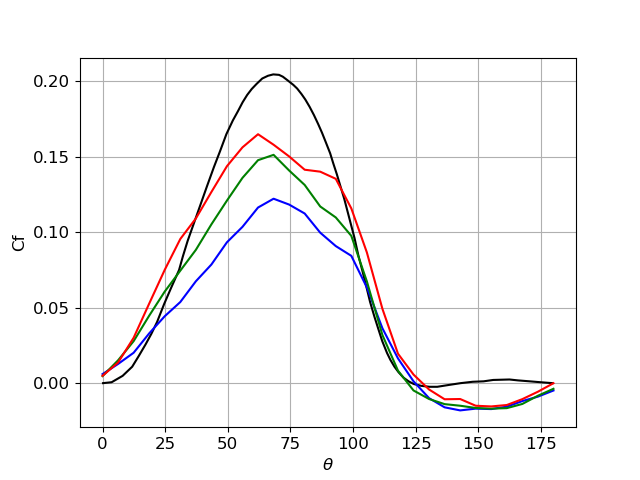} \\
(a) & (b)
\end{tabular}
\caption{Wall effects observed on the surface of the immersed sphere. The sensitivity to mesh refinement of (a) the pressure coefficient $C_p$ and (b) the friction coefficient $C_f$ is shown. Present findings are compared with results by Nagata et al. \cite{nagata_nonomura_takahashi_fukuda_2020}.}
\label{fig:sphere_bulk}
\end{figure}

At first the performance of the solver is investigated using the three grids G1, G2 and G3. Comparisons are performed for the case with $Re=300$, $M=1.2$ and adiabatic wall conditions. The drag coefficient $C_D$ is reported in Tab. \ref{tab::BulkSphere} for the three simulation considered, along with the reference results by Nagata et al.\cite{nagata_nonomura_takahashi_fukuda_2020}. One can see that the discrepancy observed for the $C_D$ is around $7\%$ for the grid G1 and it goes under $1\%$ for the grid G3. Again, one can see that the discrepancy associated with pressure component $C_{D,p}$ is always very low, lower than $2\%$. The increase of this error with mesh refinement is not problematic here, as it stays under a threshold which can reasonably be associated with uncertainty in measure. This discrepancy is interpreted here as due to the positioning of the Markers for IBM, which is probably optimal for G1 and less efficient for G3. The analysis of the component associated with the viscous friction $C_{D,v}$ shows that a significantly larger discrepancy is observed for the coarser grid G1. These findings, which are very similar to the results previously discussed for the 2D NACA profile, seem also to indicate that the present method is able to provide a relatively accurate estimation of pressure force even with coarse grids, and that the prediction of this quantity is robust. A confirmation is obtained with the analysis of the pressure coefficient $C_p$ and the friction coefficient $C_f$, which are plotted in Fig. \ref{fig:sphere_bulk} against the polar angle $\theta$. The profiles are sampled for a radius $r=D/2$ and averaged in the azimuthal direction. For $C_p$, one can see that all the numerical simulations exhibit a convincing agreement with the results by Nagata et al. [REF]. A relatively minor discrepancy is observed for the simulation using the coarse grid G1 at the stagnation point. On the other hand, results for $C_f$ do not match the reference body fitted simulations. The accuracy of the results improves with grid refinement, but a significant discrepancy is observed for all simulations.

This discrepancies could come from the used non-conservative immersed boundary method itself as described in \cite{BASKAYA2024106134} and \cite{PhysRevE.103.043302}. In hypersonic regimes, the use of non-conservative immersed boundary method (which are more flexible) induce discrepancies on the skin friction and temperature on the wall. Despite the fact that the prediction improves with mesh refinement, non-conservative methods are recommended for subsonic and supersonic regimes to keep affordable computational costs and avoid extreme refinement at the wall.

\begin{figure}

\begin{tabular}{cc}
 \includegraphics[scale=0.12]{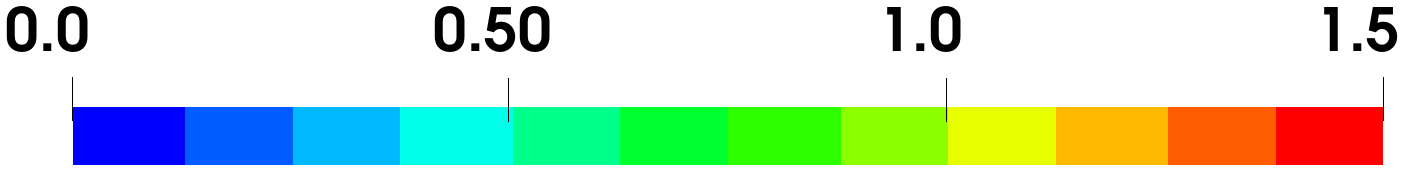} & \includegraphics[scale=0.12]{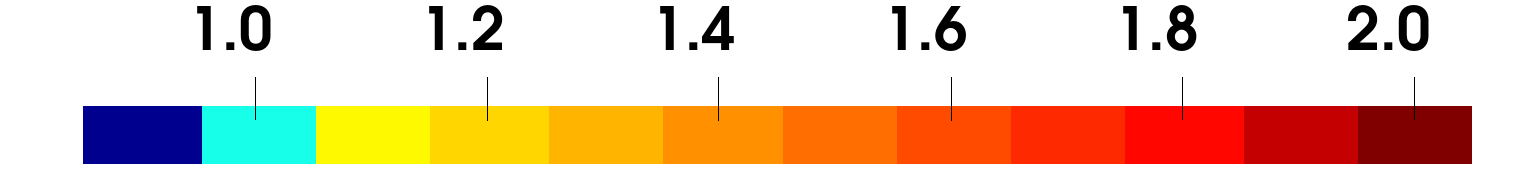} \tabularnewline
\includegraphics[scale=0.135]{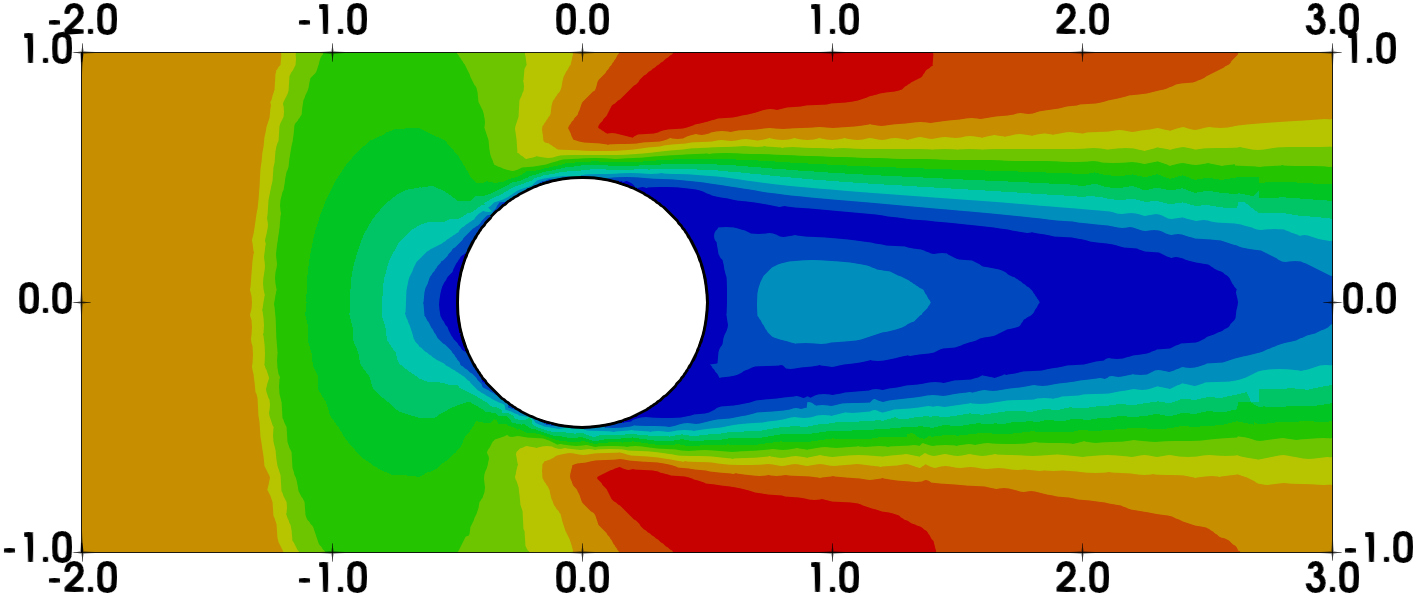} & \includegraphics[scale=0.12]{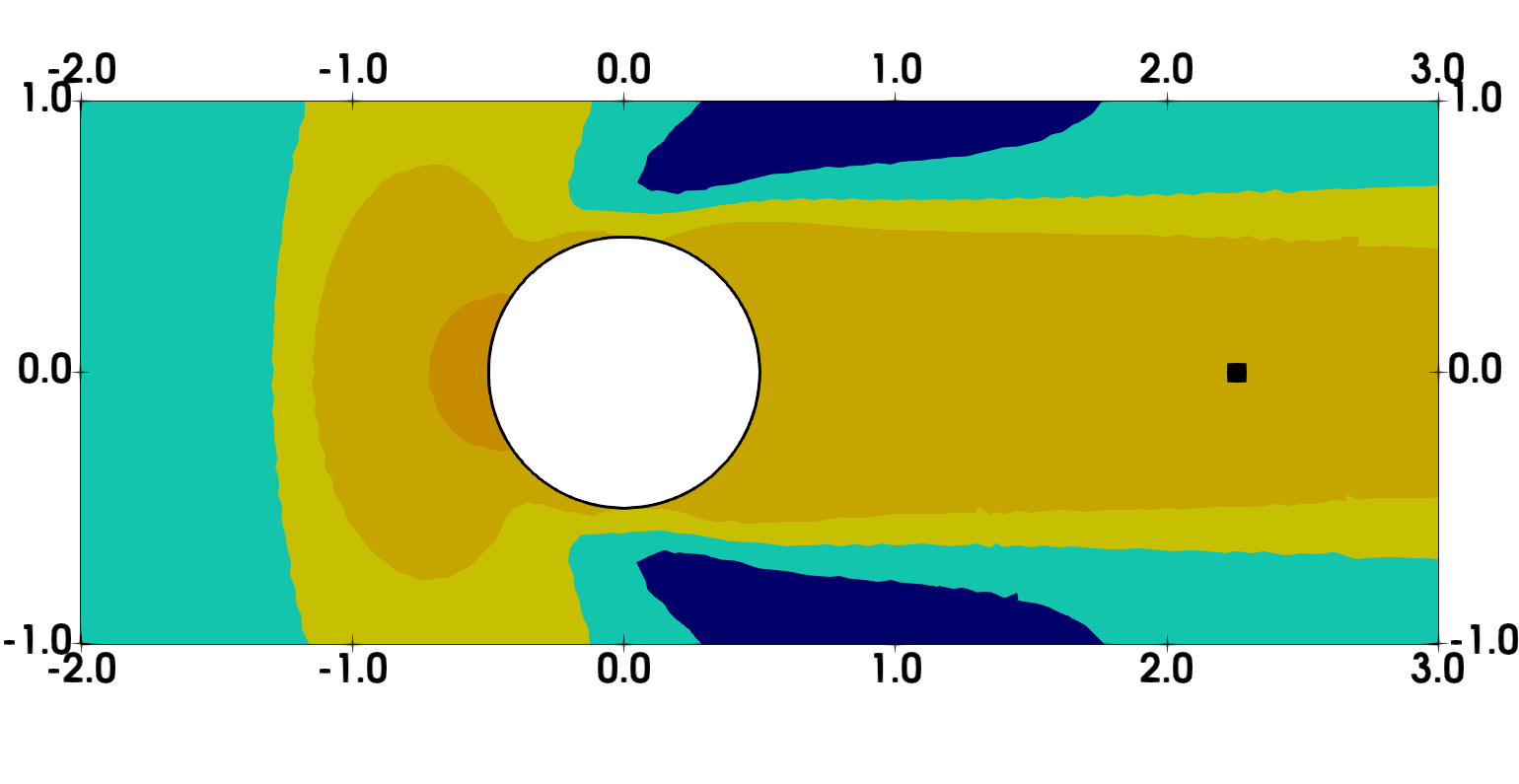} \\
\includegraphics[scale=0.135]{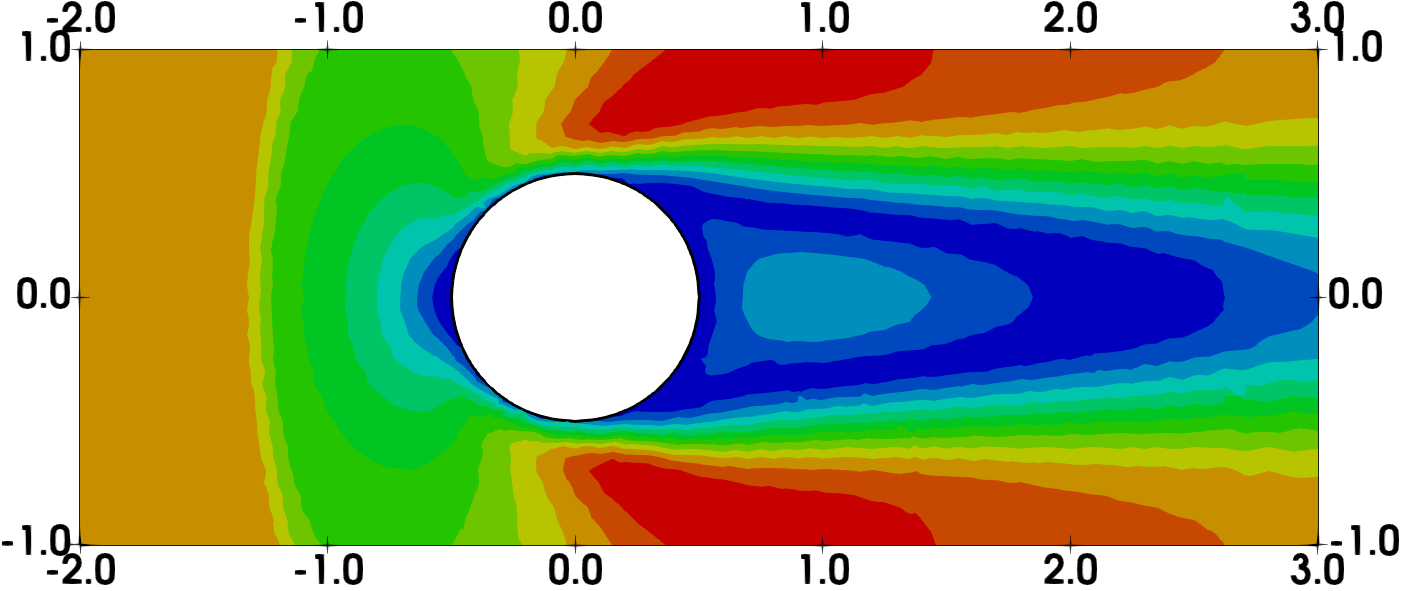} & \includegraphics[scale=0.12]{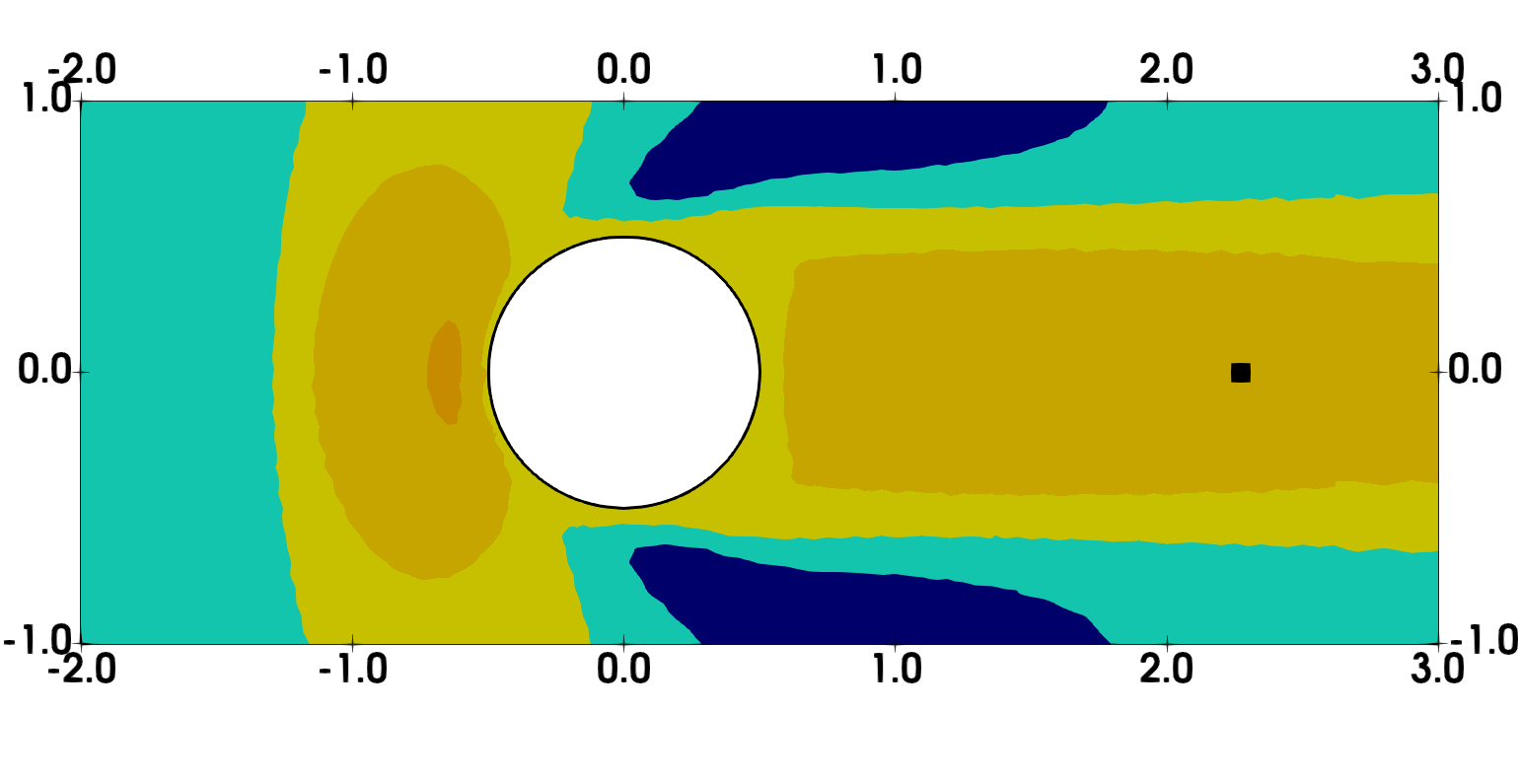} \\
\includegraphics[scale=0.135]{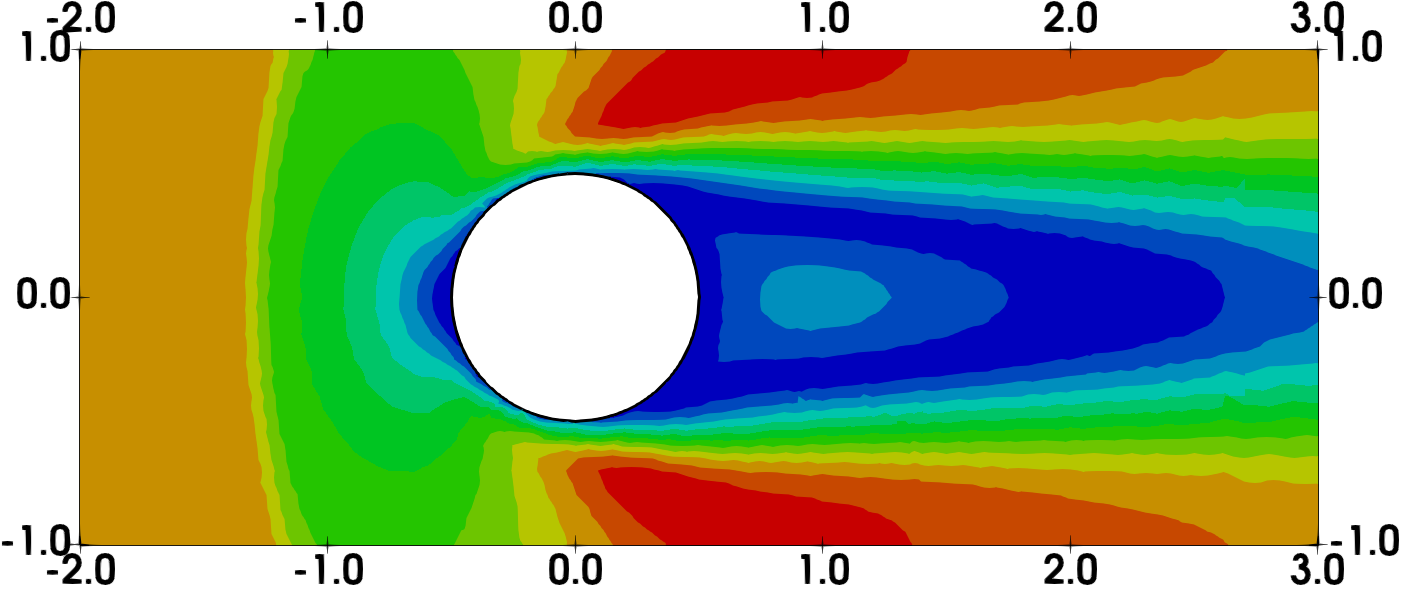} & \includegraphics[scale=0.12]{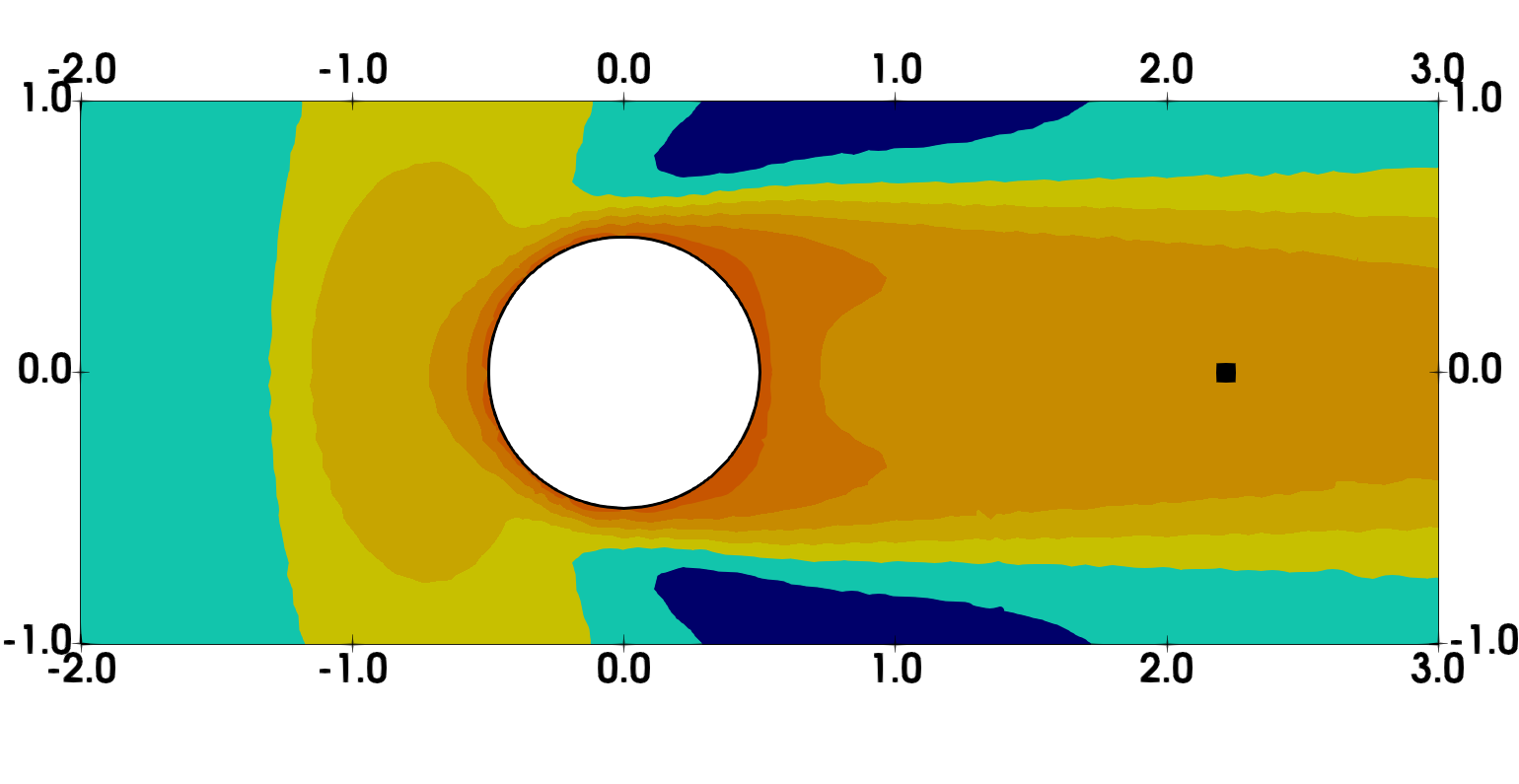} \\
\includegraphics[scale=0.135]{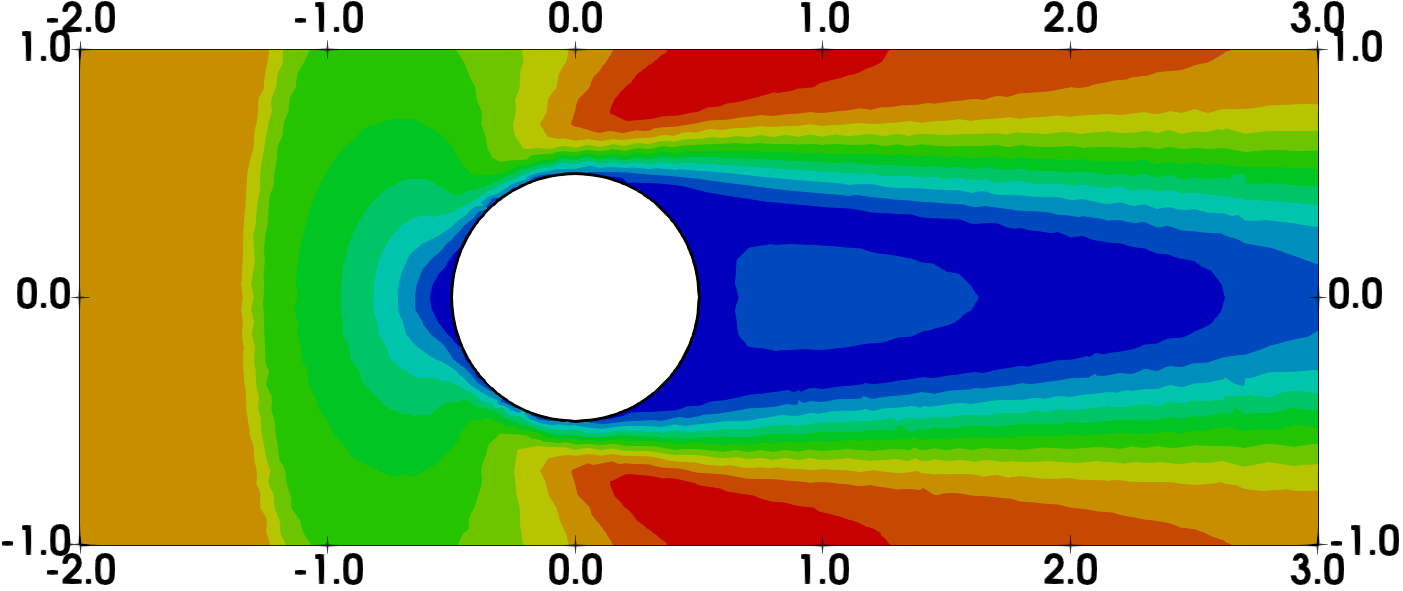} & \includegraphics[scale=0.12]{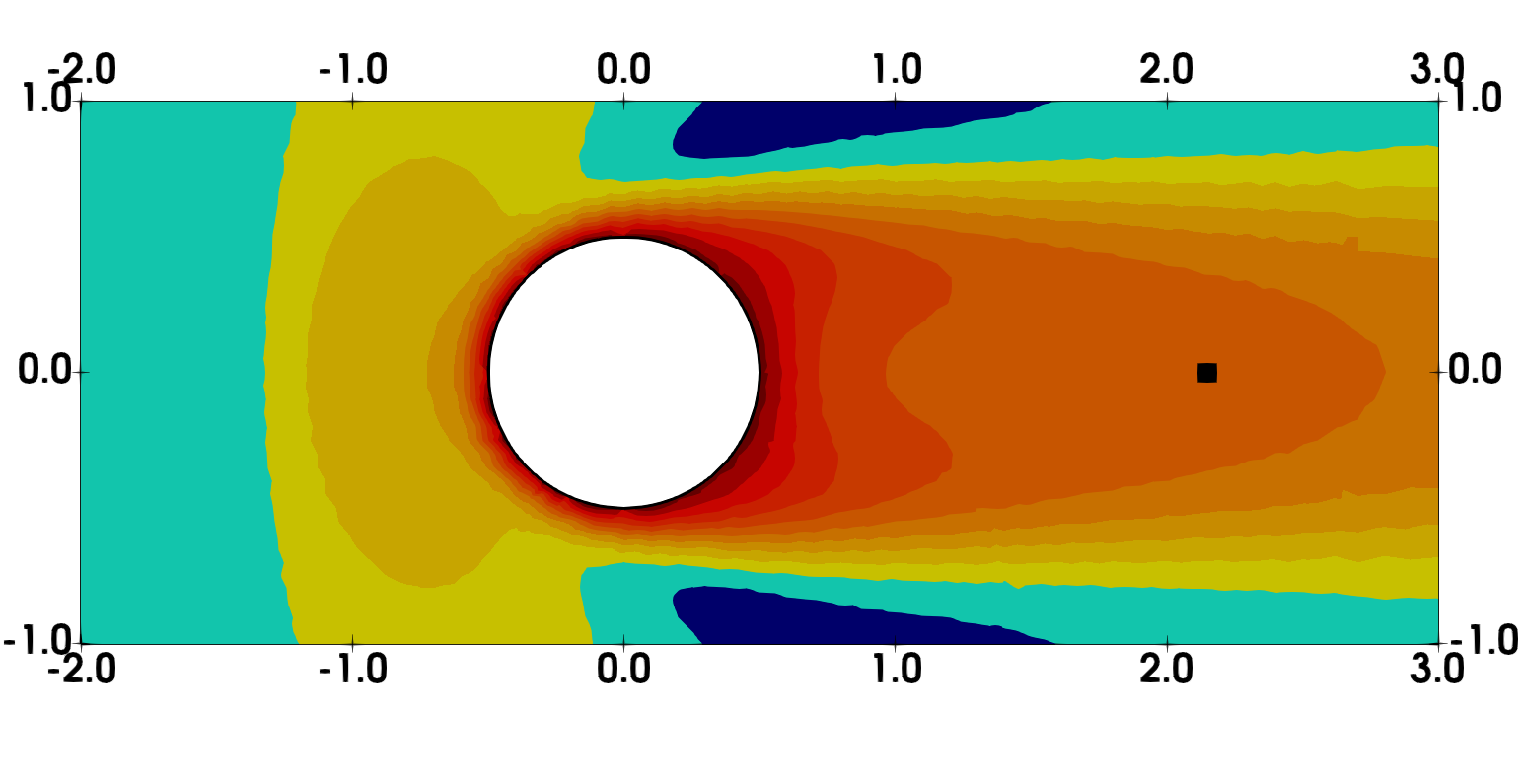} 
\end{tabular}
\caption{\label{fig::Temperature} Isocontours of (left columns) $Ma$ and (left column) the temperature for the flow around a sphere. Flow fields for each line are obtained using different wall conditions, which include adiabatic conditions (first row) as well as fixed temperature $TR=1.1$ (second row), $TR=1.5$ (third row) and $TR=2$ (fourth row). The point in the recirculation region of the figures on the right column correspond to the end of the recirculation region at the centerline.}
\end{figure}

\begin{figure}
\begin{tabular}{cc}
\includegraphics[scale=0.4]{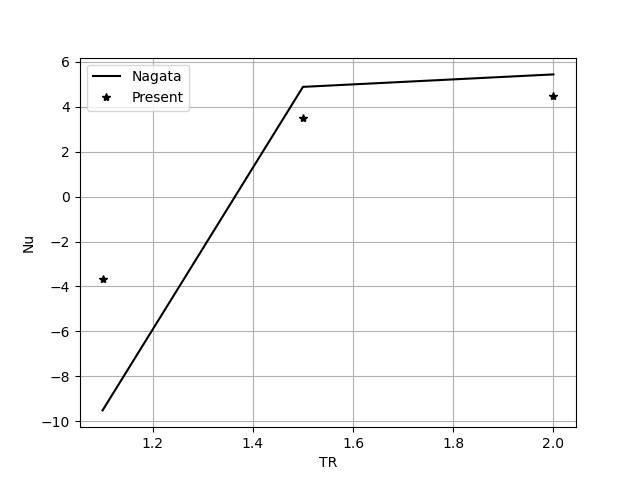} & \includegraphics[scale=0.4]{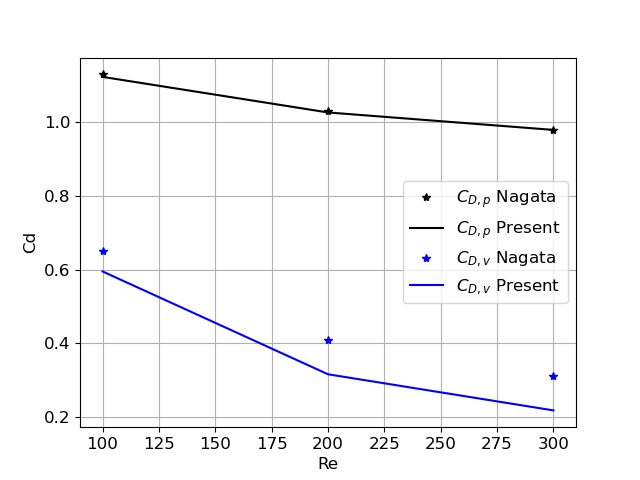} \\
(a) & (b)
\end{tabular}
\caption{\label{NUandCD} Sensitivity of the bulk quantities to variations in the conditions imposed. (a) Sensitivity of the global Nusselt number $Nu$ with the temperature imposed at the wall. (b) Sensitivity of the pressure and friction components of the drag coefficient $C_D$ to the Reynolds number $Re$.}
\label{fig:sphere_Cf}
\end{figure}

Now the usage of different thermal wall boundary conditions is analyzed. This investigation aims to assess the accuracy of the present IBM to capture different physical conditions at the wall. For this analysis, the Reynolds number is fixed at $Re=300$ and the computational grid used is G1 for every simulation. Results are presented in Fig. \ref{fig::Temperature}. One can see that the usage of different wall thermal conditions produces very different physical features, accordingly to the different physical features imposed at the wall. The analysis of the average Nusselt number $Nu$, which is shown in figure \ref{NUandCD} (a), shows significant variations with changes in the parameter $TR$. These variations are similar to the ones observed by Nagata et al. \cite{nagata_nonomura_takahashi_fukuda_2020} At last, the sensitivity of the drag coefficient $C_D$ to the Reynolds number is shown in figure \ref{NUandCD} (b). For this case, the grid G1 is used. One can see that the pressure component of $C_D$ exhibits an excellent agreement with the body fitted results for every $Re$ investigated. On the other hand, the viscous part of $C_D$ shows a larger discrepancy. These results indicate again that the pressure contribution is well captured by this IBM tool, while the viscous component needs improvement (probably in the form of constraints for the velocity gradient at the wall). However, considering the very moderate computational resources required, this tool exhibits remarkable efficiency features. 

\begin{table}
\begin{centering}
\begin{tabular}{|c|c|c|c|c|c|}
\hline 
Case & $\Delta x=\Delta y=\Delta z$ & number of cells & $Re$ & $Ma$ & Thermal BC\tabularnewline
\hline 
\hline 
\texttt{sphere\_R100\_G1\_Adiab} & $D/20$ & 350k & 100 & 1.2 & Adiabatic\tabularnewline
\hline 
\texttt{sphere\_R200\_G1\_Adiab} & $D/20$  & 350k  & 200 & 1.2 & Adiabatic \tabularnewline
\hline 
\texttt{sphere\_R300\_G1\_Adiab} & $D/20$ & 350k & 300 & 1.2 & Adiabatic \tabularnewline
\hline 
\texttt{sphere\_R300\_G2\_Adiab} & $D/40$ & 1M & 300 & 1.2 & Adiabatic \tabularnewline
\hline 
\texttt{sphere\_R300\_G3\_Adiab} & $D/80$ & 3.4M & 300 & 1.2 & Adiabatic \tabularnewline
\hline 
\texttt{sphere\_R300\_G1\_TR1.1} & $D/20$ & 350k & 300 & 1.2 & Isothermal $TR=1.1$\tabularnewline
\hline 
\texttt{sphere\_R300\_G1\_TR1.5} & $D/20$ & 350k & 300 & 1.2 & Isothermal $TR=1.5$\tabularnewline
\hline 
\texttt{sphere\_R300\_G1\_TR2} & $D/20$ & 350k & 300 & 1.2 & Isothermal $TR=2.0$\tabularnewline
\hline 
\end{tabular}
\par\end{centering}
\caption{\label{tab::DataBSphere} Characteristics of the database of numerical simulations for the 3D flow around the sphere. Details for the
grid, $Re$, $Ma$ and the thermal boundary condition are provided.}
\end{table}

\begin{table}
\begin{centering}
\begin{tabular}{|c|c|c|c|c|c|c|}
\hline 
Case & $C_{D}$ &  & $C_{D,p}$ &  & $C_{D,v}$ & \tabularnewline
\hline 
\hline 
Nagata et al. \cite{nagata_nonomura_takahashi_fukuda_2020} & 1.288 &  & 0.978 &  & 0.31 & \tabularnewline
\hline 
\texttt{sphere\_R300\_G1\_Adiab} & 1.197 & 7.07\% & 0.979 & 0.08\% & 0.22 & -6.99\%\tabularnewline
\hline 
\texttt{sphere\_R300\_G2\_Adiab} & 1.239 & 3.80\% & 0.982 & 0.31\% & 0.26 & -3.88\%\tabularnewline
\hline 
\texttt{sphere\_R300\_G3\_Adiab} & 1.3 & 0.93\% & 1.0 & 1.71\% & 0.3 & -0.78\%\tabularnewline
\hline 
\end{tabular}
\par\end{centering}
\caption{\label{tab::BulkSphere} Sensitivity of the drag coefficient to the grid used for calculations. Results from three simulations of the database are compared with reference values from Nagata et al.  \cite{nagata_nonomura_takahashi_fukuda_2020}.}
\end{table}


\clearpage

\section{Simulation of a Re-entry vehicle}
\label{sec::vehicle}

In the frame of national and international space law regulations, a large number of tools has been developed in order to treat the complexity of the physical processes involved in an atmosphere reentry of vehicles and debris. Nevertheless, important uncertainties in physical aspects of re-entry remain due to the lack of data and knowledge about the calibration of numerical models. Therefore, the development of complex solvers with coupled physics  and their improvement / validation using experimental data essential. This Section of the manuscript is devoted to such type of analysis, and in particular the IBM here developed is used to investigate a realistic test case for atmospheric reentry. In particular, the vehicle used for the IXV mission \cite{HAYARAMOS201639} is chosen for this purpose. The main objective of the IXV mission was to carry out a series of in-flight experiments and the vehicle was instrumented with numerous pressure sensors. On 11 February 2015, the IXV conducted its first 100-minute space flight, successfully completing its mission upon landing intact on the surface of the Pacific Ocean.
The spaceplane separated from the Vega launcher  and begun its reentry at 120 km altitude, travelling at a recorded speed of 7.5 km/s, identically  to a typical re-entry path of a low Earth orbit (LEO) spacecraft. This flight represents a unique means of validation of the computational tools employed for the design of such mission.The IXV flight provides extremely valuable data for code validation purposes. 

\begin{figure}
\begin{tabular}{cc}
\includegraphics[scale=0.12]{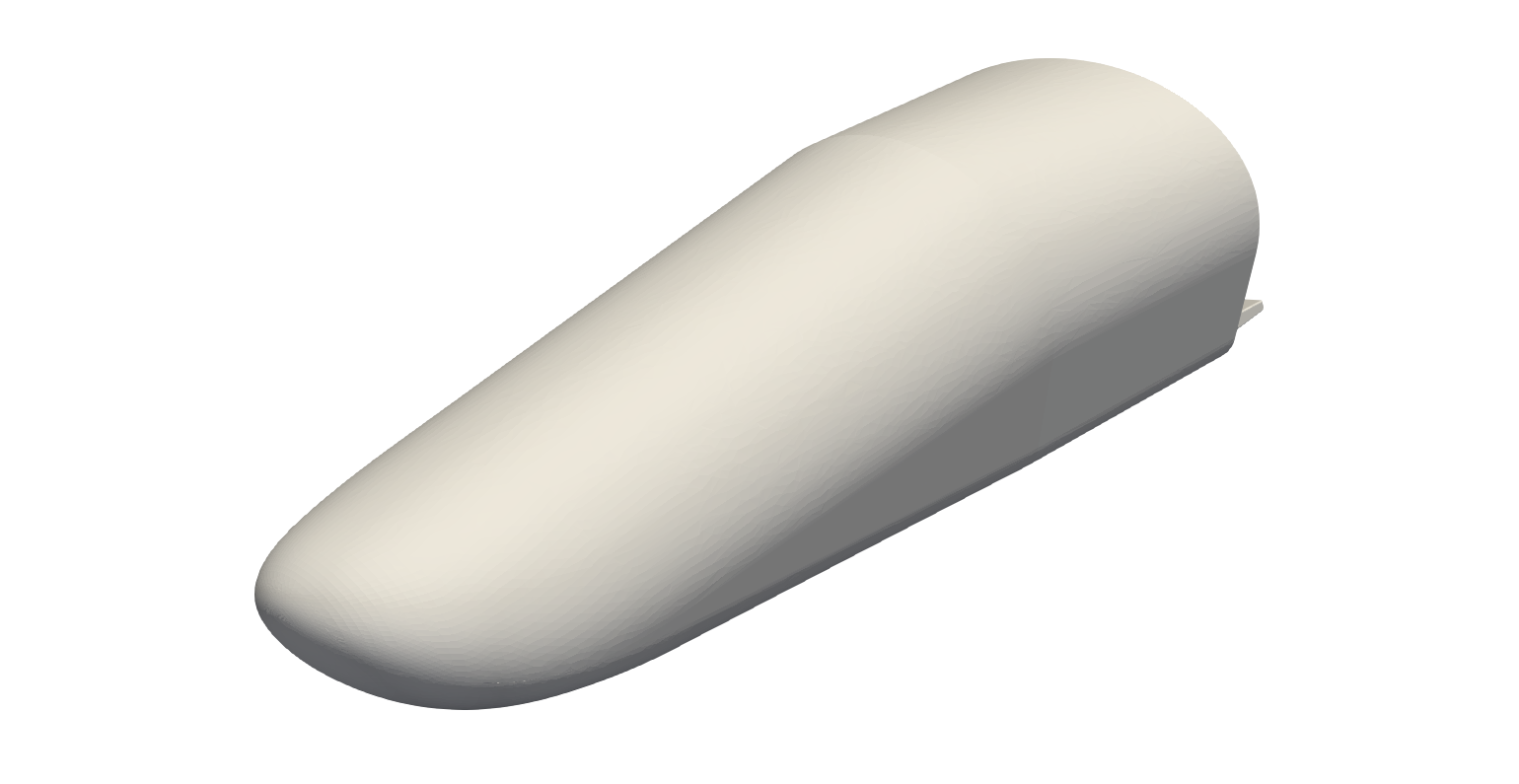} & \includegraphics[scale=0.12]{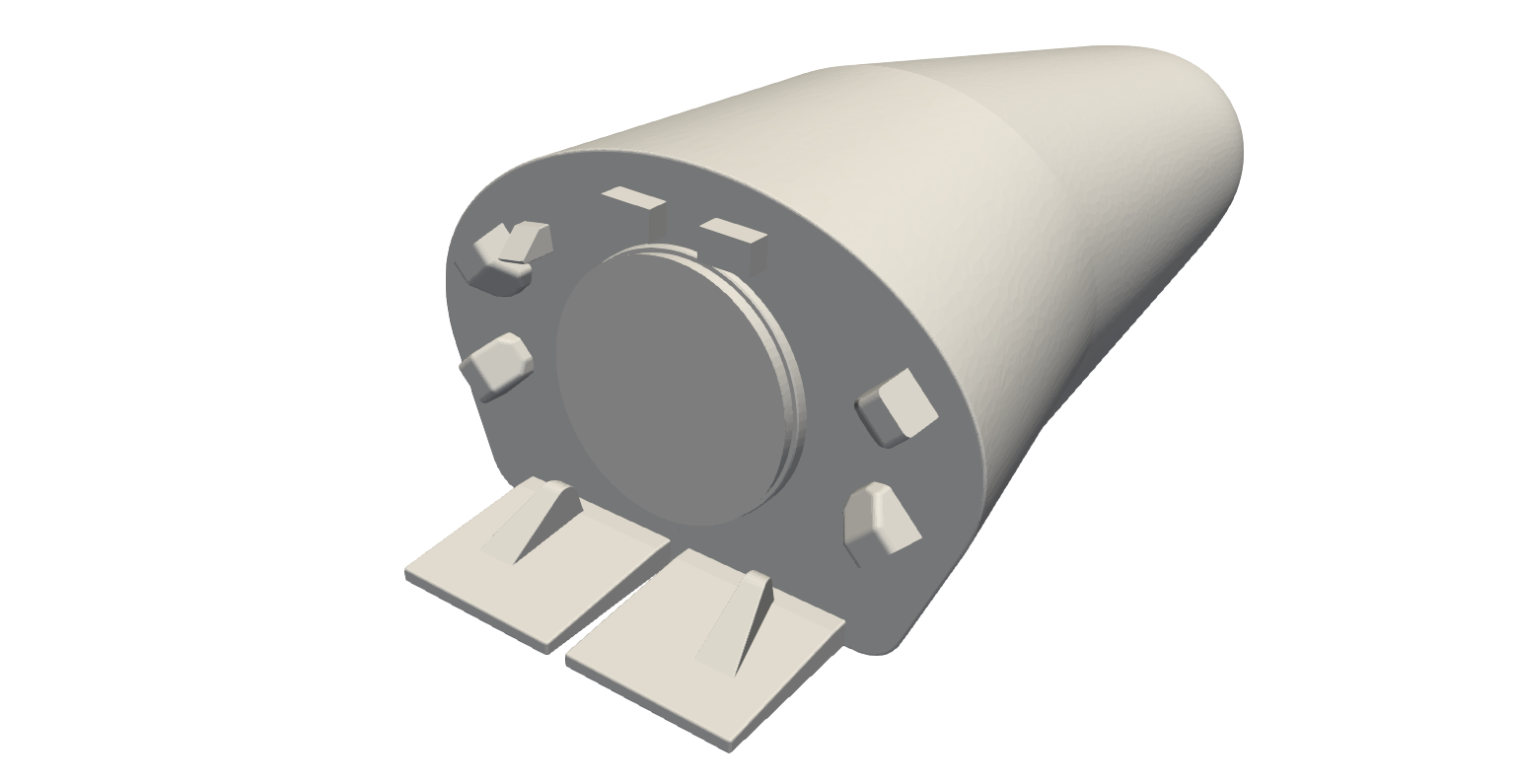} \\
Front view & Back view
\end{tabular}
\caption{\label{fig:ixv_geometry}Geometric features of the space vehicle used in the IXV mission.}
\end{figure}

\subsection{Case setup}

The set-up of the numerical IBM simulations is now presented. The aim of this analysis is to validate its capabilities to predict the latter stages of the reentry. To this end four flight configurations with different $Ma$ have been selected and calculated using the IBM solver. The geometry used for this investigation, which is  shown in Fig. \ref{fig:ixv_geometry}, includes a number of complex features. Their representation via IBM is not trivial and therefore it is a good test to investigate the performance of the model when applied to realistic cases. If we denote by $L$ the longitudinaml length of the vehicle, the physical domain investigated is about $100L$x$40L$x$40L$. The Cartesian grid used for this investigation is composed by about 2 million elements, which corresponds to a very coarse level of resolution. The cell size of the grid around the vehicle is aqual to $0.01L$. The advantage of this strategy is that multiple realizations can be performed with reduced computational demands.

\begin{table}
\begin{centering}
\begin{tabular}{|c||c|c|c|c|}
\hline 
Case & IXV\_TP1 & IXV\_TP2 & IXV\_TP3 & IXV\_TP4 \tabularnewline
\hline 
Far field Mach number & 4.27 & 3.50 & 2.76 & 1.88 \tabularnewline
\hline 
\end{tabular}
\par\end{centering}
\caption{\label{tab::configIXV} Configurations selected for the IBM simulations of the space vehicle. The configurations have been chosen because of availability of experimental / numerical reference data.}
\end{table}

\subsection{Results and discussion}

The results obtained via IBM for the four configurations reported in Tab. \ref{tab::configIXV} are now analyzed. The results are compared with experimental samples \cite{HAYARAMOS201639} and with results obtained in a previous numerical study carried out with MISTRAL-CFD solver \cite{spel2022}.  The reference simulations were performed with a structured body-fitted mesh grid, which was designed to directly resolve the active scales at the wall in the region where the boundary layer is attached. {The total number of mesh elements used for this Mistral simulation is 39k cells on the coarse mesh and 2.5M cells on the fine mesh}.

\begin{figure}
\begin{tabular}{ccc}
\includegraphics[scale=0.35]{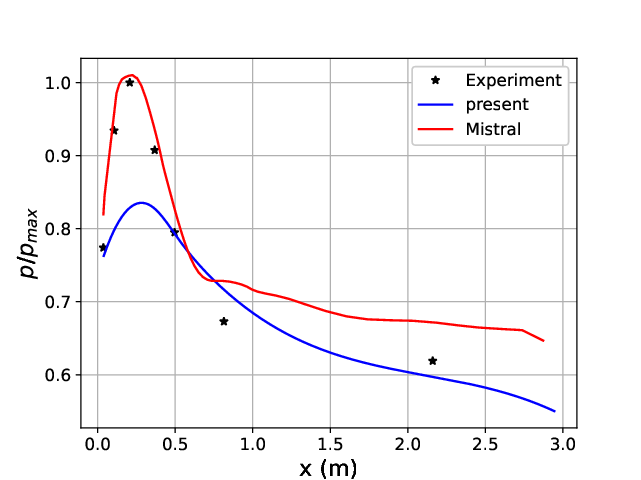} & \includegraphics[scale=0.35]{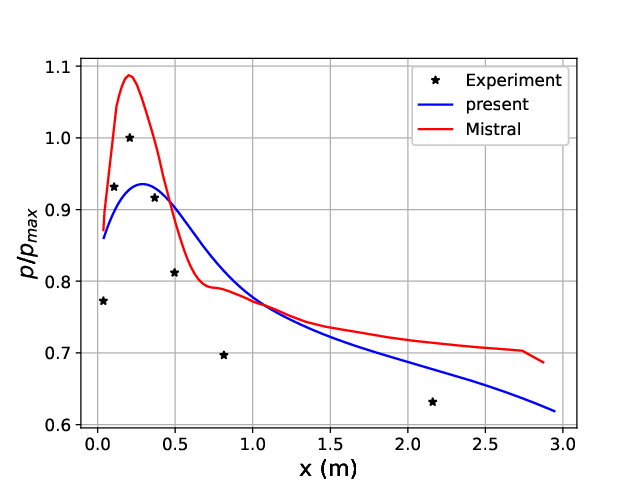} & \includegraphics[scale=0.4]{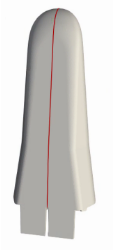} \\
(a) & (b) & \\
\includegraphics[scale=0.35]{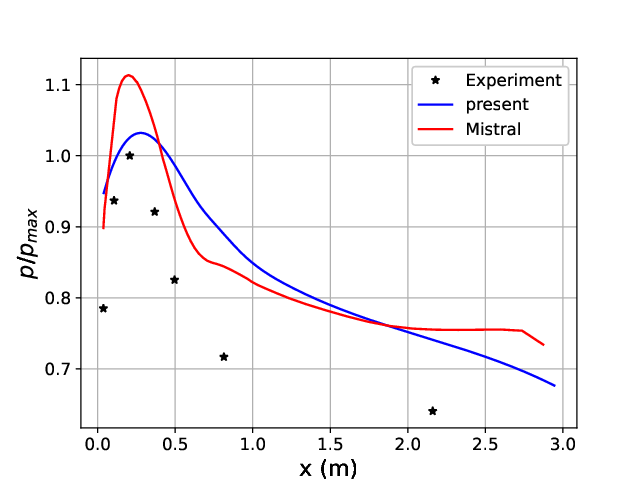} & \includegraphics[scale=0.35]{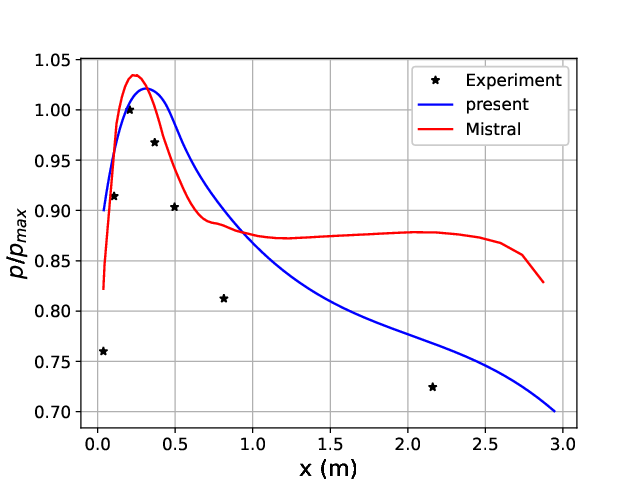} & \\
 (c) & (d) &
\end{tabular}
\caption{\label{fig:ixv_pressure} Pressure at the wall along the center line of the lower surface of the IXV vehicle at different trajectory points. The results are shown for the flight configurations (a) IXV\_TP1, (b) IXV\_TP2, (c) IXV\_TP3 and (d) IXV\_TP4}
\end{figure}

The pressure along the center line of the lower surface of the vehicle is plotted in Fig. \ref{fig:ixv_pressure} for the four trajectory points. The values are normalized by the the maximal experimental value at each trajectory point. It can be seen that pressure at the stagnation point is relatively well predicted by the present IBM simulation for the trajectory points IXV\_TP3 and IXV\_TP4 i.e. for $Ma<3$. However, this quantity is under predicted for the cases IXV\_TP1 and IXV\_TP2 with a relative error of 16\% and 7\% respectively. The profilesof the pressure away from the stagnation point is relatively similar to the prediction of MISTRAL-CFD for the three cases IXV\_TP1, IXV\_TP2 and IXV\_TP3. For the case IXV\_TP4, the pressure obtained by the IBM is even closer to experimental results than the calculation via Mistral CFD. However, the IBM calculations were performed using a significantly coarser grid, therefore similar accuracy was obtained with significantly lower computational resources when compared with the reference body-fitted simulation.

Overall, the results observed for this test case indicate what was previously observed for the airfoil and the sphere. The IBM here developed produces a reliable and precise prediction of the pressure field even when using coarse grids. For this specific test case, advancement is needed in terms of turbulence modelling. In fact, the IBM developed does not yet take into account corrections for the near-wall turbulence, in particular scaling laws to be applied to the turbulent viscosity such as wall functions \cite{Pope2000_cambridge,Wilcox2006_DCW}. 

\section{Conclusions}
\label{sec:Conclusions}

In the present manuscript a new Immersed Boundary Method for the simulation of flows for aerospace applications has been presented. This method exhibits similar features to the sharp IBM proposal by Peron et al. \cite{peron2021immersed}, but the present formalism can be integrated within implicit flow solvers which are extensively used in aerospace applications. This algorithmic formulation has been obtained integrating a Luenberger observer in the formalism of the IBM, which acts as a controller to impose the right boundary conditions at the wall. Thanks to its flexibility, both Dirichelet and Neumann conditions can be applied for the flow field of interest, permitting to represent complex wall dynamics.

The IBM algorithm, which was integrated in the opensource code OpenFOAM, has been validated via the simulation of progressively more complex test cases, namely the flow around a NACA profile, the flow around a sphere and at last a flow around a space vehicle. Sensivity analyses to the numerical set-up of the model as well as variations in the physical properties of the flow have been performed to assess the accuracy and robustness of the model. The analysis of the results indicated shared features for the three test cases. In fact, the method is able to reproduce the pressure distribution around the body even using coarse meshes. Therefore, these results highlight the very favorable accuracy versus cost feature of this method, which is conserved even when the geometry of the body investigated becomes more complex. This aspect shows potential for preliminary analyses of realistic configurations of industrail interest. The main limit of this model is associated with the representation of the velocity gradients at the wall, which significantly impact the prediction of the wall shear stress. Future works will be devoted to the inclusion of additional terms in the Luenberger observer to provide constraints to the velocity gradients. In addition, turbulence modelling in terms of wall laws will also need to be integrated in order to obtain accurate predictions for realistic applications.

\bibliographystyle{unsrt}
\bibliography{ref}

\end{document}